\documentclass{aa}
\usepackage{natbib}
\usepackage{hyperref}
\hypersetup{colorlinks=true, citecolor=blue}
\usepackage{graphicx}
\usepackage{txfonts}
\bibpunct{(}{)}{;}{a}{}{,}
\begin{document}
  \title{The IRX--$\beta$ relation on sub--galactic scales in star--forming galaxies of the {\it Herschel}\thanks{{\it Herschel} is an ESA space observatory with science instruments provided by European-led Principal Investigator consortia and with important participation from NASA.} Reference Survey}
  \author{M. Boquien\inst{1}, V. Buat\inst{1}, A. Boselli\inst{1}, M. Baes\inst{2}, G. J. Bendo\inst{3}, L. Ciesla\inst{1}, A. Cooray\inst{4}, L. Cortese\inst{5}, S. Eales\inst{6}, G. Gavazzi\inst{7}, H. L. Gomez\inst{6}, V. Lebouteiller\inst{8}, C. Pappalardo\inst{9}, M. Pohlen\inst{6}, M. W. L. Smith\inst{6}, L. Spinoglio\inst{10}}
  \authorrunning{Boquien et al.}
  \institute{Laboratoire d'Astrophysique de Marseille - LAM, Universit\'e Aix-Marseille \& CNRS, UMR7326, 38 rue F. Joliot-Curie, 13388 Marseille CEDEX 13, France \email{mederic.boquien@oamp.fr}
  \and Sterrenkundig Observatorium, Universiteit Gent, Krijgslaan 281-S9, B-9000 Gent, Belgium
  \and UK ALMA Regional Centre Node, Jodrell Bank Centre for Astrophysics, School of Physics and Astronomy, University of Manchester, Oxford Road, Manchester M13 9PL, UK
  \and Dept. of Physics \& Astronomy, University of California, Irvine, CA 92697, USA
  \and European Southern Observatory, Karl-Schwarzschild Str. 2, 85748 Garching bei M\"unchen, Germany
  \and School of Physics and Astronomy, Cardiff University, The Parade, Cardiff, CF24 3AA, UK
  \and Universit\`a degli Studi di Milano-Bicocca, Piazza della Scienza, 3, 20126, Milano, Italy
  \and Laboratoire AIM, CEA/DSM-CNRS-Universit\'e Paris Diderot DAPNIA/Service d'Astrophysique B\^at. 709, CEA-Saclay, F-91191 Gif-sur-Yvette CEDEX, France
  \and INAF-Osservatorio Astrofisico di Arcetri, Largo Enrico Fermi 5, 50125 Firenze, Italy
  \and IFSI-INAF, Via del Fosso del Cavaliere 100, 00133 Rome, Italy
  }
  \date{}
  \abstract
  {Ultraviolet and optical surveys are essential to gain insight into the processes driving galaxy formation and evolution. The rest--frame ultraviolet emission, redshifted to optical and near--infrared for high--redshift galaxies, is key to measure the cosmic star formation rate. However, the ultraviolet light is strongly reddened and absorbed by dust. In starburst galaxies, the ultraviolet colour and the attenuation are intrinsically linked, allowing to correct for dust extinction. Unfortunately, evidence has been accumulating that the relation between ultraviolet colour and attenuation is different for normal star--forming galaxies when compared to starburst galaxies.}
  {It is still not understood why star--forming galaxies deviate from the ultraviolet colour--attenuation relation of starburst galaxies. Previous work and models hint that the role of the shape of the attenuation curve and the age of stellar populations have an important role. In this paper we aim at understanding the fundamental reasons to explain this deviation.}
  {We have used the CIGALE spectral energy distribution fitting code to model the far ultraviolet to the far infrared emission of a set of 7 reasonably face--on spiral galaxies from the Herschel Reference Survey on a pixel--by--pixel basis. We have explored the influence of a wide range of physical parameters to quantify their influence and impact on the accurate determination of the attenuation from the ultraviolet colour, and why normal galaxies do not follow the same relation as starburst galaxies.}
  {We have found that the deviation from the starburst relation can be best explained by intrinsic ultraviolet colour differences between different regions in galaxies. Variations in the shape of the attenuation curve can also play a secondary role. Standard age estimators of the stellar populations such as the $D4000$ index or the birthrate parameter prove to be poor predictors of the intrinsic ultraviolet colour. These results are also retrieved on a sample of 58 spiral galaxies drawn from the Herschel Reference Survey sample when considering their integrated fluxes.
  }
  {When correcting the emission of normal star--forming galaxies for the attenuation, it is crucial to take into account possible variations in the intrinsic ultraviolet colour of the stellar populations as well as variations of the shape of the attenuation curve.
  }
  \maketitle

\section{Introduction\label{sec:introduction}}

Over the last decade, large, deep surveys of high redshift galaxies have been dedicated to gaining insight into the physical processes at play in the formation and evolution of galaxies across the Universe. The major process in the transformation of baryonic matter is star formation. It converts the local gas reservoir into heavy elements that are ejected from star--forming regions by way of feedback, seeding the interstellar medium and the intergalactic medium with metals. One of the main constraints on cosmological models is the so--called cosmic star formation rate (SFR) density, which has been intensely studied since the seminal work of \cite{madau1996a} both observationally \citep[e.g. ][]{steidel1999a,hopkins2004a,perez2005a,schiminovich2005a,bouwens2009a,magnelli2009a,reddy2009a,rodighiero2010a,vanderburg2010a,magnelli2011a} and theoretically \citep[e.g. ][]{kitzbichler2007a,dave2011a}.

To measure the SFR across the Universe, the ultraviolet (UV) is theoretically the wavelength domain of choice for high--redshift galaxies. Indeed, rest--frame UV radiation is redshifted into optical and near--infrared bands which are easily accessible with broadband observations from the ground. UV is a direct tracer of star formation as it is sensitive to the photospheric emission of massive stars. Unfortunately the presence of dust affects its effectiveness: it reddens the UV--optical spectral energy distribution (SED) as it absorbs energetic radiation that is re--emitted in the mid-- to far--infrared (IR), and it is also affected by scattering out and into the line of sight. Following the usual convention, extinction encompasses absorption and scattering out of the line of sight while attenuation also takes into account the scattering into the line of sight. Correcting the UV emission for the attenuation is therefore a crucial requirement to estimate the actual SFR of galaxies.

A powerful way to correct for the attenuation is to combine attenuation sensitive star formation tracers with the IR emission \citep[e.g. ][]{calzetti2007a,leroy2008a,kennicutt2009a,hao2011a}. Unfortunately IR data are not necessarily available and often still not at a sufficient depth and/or resolution, even with {\it Herschel}. In this context, one of the most commonly used methods to correct for the attenuation is the so--called IRX--$\beta$ relation \citep{calzetti1994a,meurer1999a,calzetti2000a}. This method links the observed UV slope $\beta$ to $IRX$, a measure of the attenuation:
\begin{equation}
IRX\equiv\log\left(L_{dust}/L_{FUV}\right),\label{eqn:IRX}
\end{equation}
with $L_{dust}$ the total luminosity of the dust, and $L_{FUV}$ the far UV luminosity computed as $\nu L_\nu\left(\lambda\right)$ with $\nu$ the frequency and $L_\nu\left(\lambda\right)$ the monochromatic luminosity per unit frequency at the wavelength $\lambda$. While this relation provides us with accurate estimates of the attenuation for starburst galaxies \citep{meurer1999a}, it fails, sometimes considerably, for galaxies forming stars at a lower rate. The fundamental reason for this deviation from the starburst law is still poorly understood. Various studies hint at the possible role of the age of the stellar populations and/or variations of the shape of the attenuation curve \citep{bell2002a,kong2004a,burgarella2005a,seibert2005a,cortese2006b,gildepaz2007a,johnson2007a,panuzzo2007a,cortese2008a,boquien2009b,munoz2009a,wijesinghe2011a}. This is a severe problem as surveys go ever deeper and probe the fainter end of the luminosity function which is populated by galaxies that can strongly deviate from the starburst law. The now routine detection of normal star--forming galaxies at high redshift shows that there is urgency to understand what is the physical origin of this deviation.

Observations over a broad wavelength range is key to understanding the physical processes at play in star--forming galaxies. Indeed, the energy output of young stellar populations is dominated by the photospheric emission of short--lived, blue, massive stars that emit in the UV. These stars also ionise the surrounding gas which recombines emitting lines that can dominate the flux in optical bands \citep{anders2003a}. At the same time the optical is also a combination of young and old stellar populations. The relative weight of the emission from these two populations changes with wavelength with younger populations dominating at short wavelengths whereas the evolved population dominating at longer wavelengths, peaking in the near--IR. To disentangle these populations a full sampling in the UV, optical, and in the near--IR is needed. These bands are however sensitive to the dust which absorbs the emission of stars and re--emits the energy at longer wavelength in the mid-- and far--IR. The emission in the mid--IR arises principally from hot dust stochastically heated and emission bands from large complex molecules such as PAH. Conversely the far--IR is dominated mainly by warm dust ($\sim50$~K, under $\sim100~\mu$m) and cold dust ($\sim20$~K, beyond $\sim100~\mu$m), as can be seen in M33 for instance \citep{kramer2010a}. The emission of the warm dust in the far--infrared is particularly important to constrain the fraction of extinguished star formation as it dominates the energy budget. Understanding why star--forming galaxies deviate from the starburst relation therefore requires a multi--wavelength data set from the UV to the FIR.

In this article we investigate, disentangle, and quantify the role of a large set of physical parameters to explain the deviation of star--forming galaxies from the starburst IRX--$\beta$ relation. To do so, we study the IRX--$\beta$ relation using FUV to FIR data for seven resolved galaxies from the {\it Herschel} Reference Survey \citep[HRS,][]{boselli2010a}. 

In Sec.~\ref{sec:data} we explain how we selected the galaxies to be studied and how we processed the multi--wavelength data. Each data point has been modelled using CIGALE \citep[Code Investigating GALaxy Emission,][]{noll2009a} to reproduce their SED from the FUV to the FIR as presented in Sec.~\ref{sec:SED-fitting}. In Sec.~\ref{sec:discussion} we analyse the direct estimate of a large number of physical parameters allowing us to perform a detailed study of the relation between these parameters, $\beta$, and $IRX$, and in particular to constrain the effect of the age of the stellar population and the shape of the attenuation curves. Finally, we conclude in Sec.~\ref{sec:conclusion}

\section{Sample and data\label{sec:data}}

\subsection{Sample selection}

The sample selection is guided by the core question of this paper: why and how do normal spiral galaxies forming at most a few M$_{\sun}$~yr$^{-1}$ of stars deviate from the IRX--$\beta$ relation that has been derived for starburst galaxies? To answer this question we must be able to disentangle in a resolved way the emission from the different components of the galaxy: the young and evolved stellar populations, the nebular emission, and the dust. This yields constraints on the required bands and on the resolution. In addition to ensure that this study is not affected by systematic variations in data processing we require that at each wavelength all galaxies are observed by the same instrument.

While high resolution observations are routinely available  to resolve in detail nearby star--forming galaxies in the UV, optical, and near--IR, the FIR is more challenging. The recent advent of the {\it Herschel} Space Observatory \citep{pilbratt2010a} now provides us with the sufficient resolution to resolve in detail the emission of the dust in spiral galaxies at longer wavelengths. The availability of {\it Herschel} data is therefore an absolute requirement and fundamental to this study. The largest sample of nearby galaxies has been observed by {\it Herschel} in the context of the HRS, which observed 323 galaxies in a range of environments. The HRS is a K band selected, quasi--volume limited survey between 15~Mpc and 25~Mpc. In addition, to limit the foreground contamination by Galactic cirrus, the sample is constituted of galaxies at a high Galactic latitude ($|b|>55^\circ$).

We have selected a sample of 7 galaxies (5 in the Virgo cluster) from the 323 HRS galaxies with the following criteria:
\begin{enumerate}
 \item The galaxies are normal star--forming spirals that do not have strong active nuclei that could strongly affect the emission in the UV and/or in the IR, which would induce a bias on the results.
 \item To perform a pixel--by--pixel analysis, the structure of the galaxies must be resolved at the coarsest band resolution. The reason is two--fold. First a good spatial resolution is needed to separate the various morphological components (arm and interarm regions, bulge, etc.) of each galaxy and therefore not be only dominated by the brightest regions. Then, as inclination increases, different wavelengths probe increasingly different regions in the galaxy as UV becomes optically thick more rapidly than longer wavelengths, adding much complexity to the modelling. To constrain the FIR emission while preserving the highest resolution possible, we have decided to drop the SPIRE 500~$\mu$m band, which has little consequence on the measure of the dust luminosity. To ensure galaxies are sufficiently resolved we have manually examined the images of all HRS galaxies up to 350~$\mu$m. The selection has been made on a case--by--case basis from this examination.
 \item To disentangle the emission from various components, we have selected only galaxies that have been observed in FUV, NUV, $u'$, $g'$, $r'$, $i'$, $z'$, J, H, Ks, MIPS 70~$\mu$m, SPIRE 250~$\mu$m, and SPIRE 350~$\mu$m. The MIPS 160~$\mu$m band was not considered due to its coarse resolution similar to that of the SPIRE 500~$\mu$m band. For the FUV and NUV bands, we have used the GALEX \cite[GALaxy Evolution eXplorer,][]{martin2005a} Nearby Galaxies Survey, Medium Imaging Survey, and Guest Investigator data while excluding galaxies only observed by the All--sky Imaging Survey as the data are too shallow to reach our goals. The optical data from $u'$ to $z'$ have been obtained from the SDSS \citep[Sloan Digital Sky Survey,][]{abazajian2009a}. Near--IR J, H, and Ks bands have been acquired by 2MASS \citep[2 Micron All Sky Survey,][]{skrutskie2006a}. Some of the galaxies have deeper near--IR images obtained by our team, but to prevent differences in calibration or data processing from introducing a bias, we have decided to use 2MASS data for all galaxies. Spitzer/MIPS \citep{rieke2004a} and {\it Herschel}/SPIRE \citep{griffin2010a} data have been obtained by our team \citep{bendo2012a,ciesla2012a}
\end{enumerate}
The detailed list of galaxies and their main parameters are presented in Table~\ref{tab:sample}.
\begin{table*}
\caption{Selected sample. Data provided by \cite{boselli2010a} and \cite{ciesla2012a}.}
\label{tab:sample}
\centering
\begin{tabular}{c c c c c c c c}
 \hline\hline
NGC \# & HRS \# & $\alpha$ & $\delta$ & Type & K                         & Radius & Distance\\
       &        &(J2000)   &(J2000)   &      & (Vega mag)&(\arcsec)&(Mpc)\\
\hline
3485 & 33  & 11 00 02.38 & +14 50 29.7 & Sb(r)b & 9.46 & 88 & 20.5\\
4254 & 102 & 12 18 49.63 & +14 24 59.4 & Sa(s)c & 6.93 & 258 & 17.0\\
4321 & 122 & 12 22 54.90 & +15 49 20.6 & SAB(s)bc; LINER; HII & 6.59 & 383 & 17.0\\
4519 & 196 & 12 33 30.25 & +08 39 17.1 & SB(rs)d & 9.56 & 151 & 17.0\\
4535 & 204 & 12 34 20.31 & +08 11 51.9 & SAB(s)c;HII & 7.38 & 270 & 17.0 \\
4536 & 205 & 12 34 27.13 & +02 11 16.4 & SAB(rs)bc;HII;Sbrst & 7.52 & 304 & 17.0\\
5668 & 320 & 14 33 24.34 & +04 27 01.6 & SA(s)d & 11.71 & 152 & 22.6\\
\hline
\end{tabular}
\end{table*}
Data used from the FUV to the FIR are detailed in Table~\ref{tab:data}.
\begin{table}
\caption{Multiwavelength data}
\label{tab:data}
\centering
\begin{tabular}{c c c}
\hline\hline 
Bands & Wavelength & Instrument/Survey\\
\hline
FUV            &151.6~nm& GALEX\\
NUV            &226.7~nm& "\\
$u'$             &354.3~nm& SDSS\\
$g'$             &477.0~nm& "\\
$r'$             &623.1~nm& "\\
$i'$             &762.5~nm& "\\
$z'$             &912.4~nm& "\\
J              &1.235~$\mu$m& 2MASS\\
H              &1.662~$\mu$m& "\\
Ks              &2.159~$\mu$m& "\\
MIPS 70        &70~$\mu$m& Spitzer/MIPS\\
SPIRE 250      &250~$\mu$m& {\it Herschel}/SPIRE\\
SPIRE 350      &350~$\mu$m& "\\
\hline
\end{tabular}
\end{table}

\subsection{Data processing}

To constrain the physical properties of the galaxies in a resolved way, a pixel--by--pixel SED from the FUV to the FIR has to be assembled. To do so, we have processed and convolved the data in a similar way as in \cite{boquien2011a}. The main steps are described hereafter.

\begin{enumerate}
\item Objects unrelated to the target galaxy such as foreground stars or background galaxies can be an important source of contamination when convolved to lower resolution. To limit this we have manually removed the brightest sources in the UV, the optical, and near--IR, using the {\sc imedit} procedure in {\sc iraf}.
\item For data in the UV, optical, and near--IR domains, galactic foreground extinction was corrected assuming $R_V=3.1$ with a \cite{cardelli1989a} extinction curve including the \cite{odonnell1994a} update. The differential extinction $E(B-V)$ for each galaxy was obtained from NASA/IPAC Galactic Dust Extinction Service using the reddening maps of \cite{schlegel1998a}.
\item Processing all images to a similar point spread function (PSF) is crucial to study the SED in a pixel--by--pixel way without being affected by resolution effects. To retain as many resolution elements as possible while keeping strong constraints on the emission of the dust, we have convolved the data to the PSF of the SPIRE 350~$\mu$m band (24\arcsec). We have used the large set of convolution kernels presented by \cite{aniano2011a} allowing us to degrade GALEX UV, optical/near--IR data as well as Spitzer MIPS data to the SPIRE PSF.
\item To allow for a direct pixel--by--pixel comparison, all images need to be projected on the same grid. For each galaxy we created a reference image centred on the coordinates of the galaxy, with a pixel size of 8\arcsec\  ($>659$~pc), similar to that of the SPIRE 350~$\mu$m image. The impact of the choice of the pixel size will be discussed in Sec.~\ref{ssec:pixel-size}. This allows us to Nyquist sample the PSF, making it easier to distinguish the structures. The size of the image is taken to be twice the size of the circular aperture used for the integrated photometry presented by \cite{ciesla2012a} to ensure the presence of enough sky background. All images were then registered on this reference image using the {\sc wregister} procedure in {\sc iraf}.
\item For each band we subtract the background which is taken as the median of the pixels in an annulus with an inner radius of 1.1 times the size of the photometry aperture and with a width of 1\arcmin. The photometry aperture is taken as 1.4 times the optical radius as defined by \cite{ciesla2012a}. 
\item Finally, to eliminate pixels whose physical connection to the galaxy is uncertain and those that are too faint, we only select pixels encompassed by the aperture defined by \cite{ciesla2012a}, and that are detected in all bands at a 3--$\sigma$ level considering only the uncertainties on the background level.
\end{enumerate}

\begin{figure*}[!htbp]
\centering
\includegraphics[height=0.93\textheight]{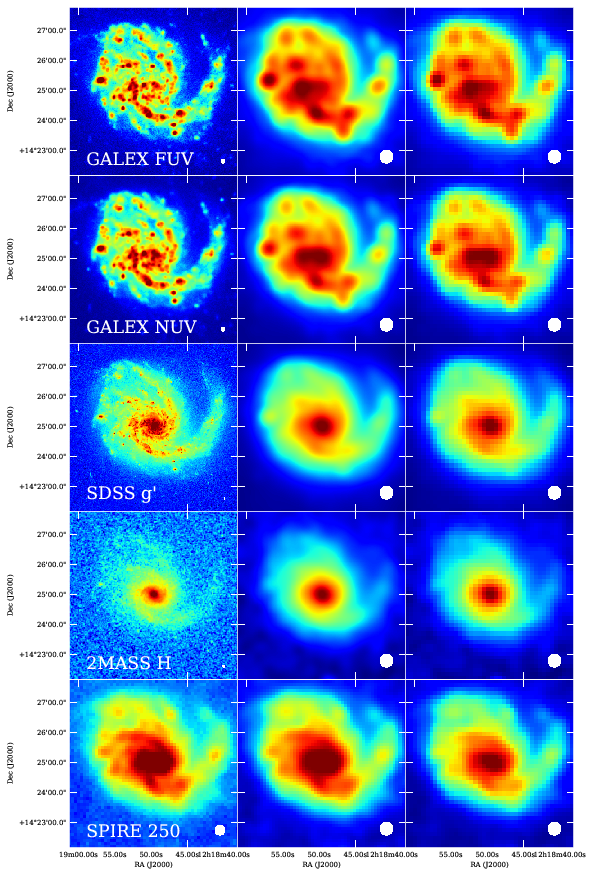}
\caption{Images of NGC~4254 (M99) in FUV, NUV, $g'$, H, and SPIRE 250~$\mu$m (from top to bottom) at different points in data processing: original images (left), original images convolved to the resolution of the SPIRE~350~$\mu$m band (centre), and final convolved images registered to the same reference frame with a pixel size of 8\arcsec\ (right). The beam size is indicated on the bottom--right hand of each panel.}
\label{fig:data-process}
\end{figure*}

An example of successive steps of the data processing from the FUV to the FIR is presented for NGC~4254 (M99) in Fig.~\ref{fig:data-process}.

\section{Spectral energy distribution fitting\label{sec:SED-fitting}}

To extract the physical parameters from the observations, we model the SED for a large set of parameters and fit these SED from the UV to the FIR. We discuss the fitting procedure including the input and output parameters hereafter.

\subsection{CIGALE}

CIGALE is a recent code to fit the SED of galaxies simultaneously from the UV to the FIR developed by our team. It has been used with success on nearby galaxies \citep{noll2009a,buat2011a} and high redshift objects \citep{giovannoli2011a,buat2011b,burgarella2011a} bringing new constraints on the physical properties of galaxies and on the attenuation laws at play.

Briefly, CIGALE models the SED of galaxies by combining several old and young stellar populations based on \cite{maraston2005a}, including nebular emission. The UV--optical domain is reddened by an attenuation law modified from the \cite{calzetti1994a,calzetti2000a} curve. This energy is re--emitted in the mid-- and far--IR which is fitted by the \cite{dale2002a} templates. The absorption and the dust emission are constrained through an energy balance.

One of the main challenges in model fitting is evaluating the uncertainties on the main parameters. One method consists in finding sets of good fits by examining the $n$--dimensions $\chi^2$ cube (in case of a standard $\chi^2$ minimisation), with $n$ the number of input parameters in the model. Indeed, even if the best fit (that is the one that minimises the $\chi^2$ value) indicates the most likely parameters, an arbitrarily large number of models can also provide reasonable fits. A range of parameters can then be determined to be acceptable. Such a method is applied in \cite{boquien2010c} for instance. With an increasingly larger volume probed by input parameters this method becomes impracticable. CIGALE, rather than only using a simple $\chi^2$ minimisation, derives the properties of a galaxy and the associated uncertainties by analysing the probability distribution function for each parameter \citep{walcher2008a,noll2009a}. This method used by CIGALE to estimate parameters and their uncertainties is described in \cite{kauffmann2003a}. The derivation of the uncertainties is presented and discussed in detail in \cite{noll2009a}.

\subsection{Input and output parameters}

The choice of the parameters are critical. We describe here the parameters involved in the modelling with CIGALE, their range, and the accuracy of the modelling. A grid of models is made varying parameters determining the stellar populations (Sec.~\ref{sssec:SF-par}) and the conversion of UV--optical photons to the IR (Sec.~\ref{sssec:ext-par}). For each model of this grid the value of the $\chi^2$ is computed and the values of the parameters are derived from the probability distribution function as described above. 

\subsubsection{Star formation parameters\label{sssec:SF-par}}

The stellar population in galaxies is made of many generations, due to a complex star formation history (SFH). However, the relative weight of old generations is increasingly smaller in the UV--optical domain as they age. This creates strong degeneracies in the SFH which is particularly difficult to constrain in detail. While complex SFH models give good fits, simpler models also give excellent fits, not allowing us to favour one model over the other. It is therefore reasonable to model the SED using 2 different bursts of star formation representing the older and the younger stellar populations. The old stellar population is modelled with an exponentially decreasing burst over $t_1=13$~Gyr with an e--folding time $\tau_1$ ranging from 3~Gyr to 7~Gyr by steps of 1~Gyr, to consider the smooth, long term secular evolution of galaxies. On top of this burst we consider a second burst to model the young stars formed during the latest star formation episode. It has been shown on entire galaxies that a constant star formation rate for this episode can reproduce convincingly the observed SED \citep{buat2011a,giovannoli2011a} as the various, individual, short bursts of star formation are averaged over the entire disk of a galaxy. In the case of resolved studies, there are fewer individual star--forming regions averaged over in sub--regions in galaxies. The assumption of a constant SFR loses its validity as the SFH of each individual star--forming region becomes more important in determining the global SED of the galaxy sub--region. To take into account the rapid variation of the SFR on small spatial scales, we have chosen to use an exponentially decreasing SFH for the young stellar population. Box--like models have also been tested and yield nearly identical results. The age $t_2$ of this burst as well as the e--folding time $\tau_2$ are left as open parameters with $2\le t_2 \le 200$~Myr and $1\le \tau_2 \le 100$~Myr, each of these parameters being logarithmically spaced with a total of 7 values each. The relative stellar mass of the young and the old populations, $f_{ySP}$, is logarithmically sampled with 6 values in the range $0.001\le f_{ySP} \le 0.316$.

\subsubsection{UV--optical attenuation and infrared emission\label{sssec:ext-par}}

The presence of dust strongly affects the SED in the UV and the optical, absorbing energetic photons to re--emit the energy in the IR. The distribution of the dust can strongly affect the shape of the effective dust attenuation curve \citep{charlot2000a,witt2000a,panuzzo2007a}. In the case of a starburst, the slope is particularly shallow, the attenuation slowly increasing with frequency \citep{calzetti1994a,calzetti2000a}. Conversely in the Small and Large Magellanic clouds the extinction curve is much steeper \citep{gordon2003a}.
In addition to intrinsic variations in attenuation laws, it is also well--known that young star--forming regions are more dusty than older ones and are therefore more extinguished. In the case of a starburst galaxy the young stellar population has an attenuation which is a factor $\sim2$ or more higher than for the old stellar population \citep{calzetti1994a,wild2011a}. Finally, as the UV and optical emission is absorbed and scattered, some radiation transfer effects could become important. At a minimal distance of 17~Mpc, a pixel size of 8\arcsec\ represents a physical size larger than 659~pc (1978~pc at the scale of the PSF). That way radiation transfer effects between adjacent regions should remain limited, though this strongly depends on the relative distribution of the dust and the stars.

Changes in the attenuation laws and the presence of a UV bump can affect the location of the data points and the models in IRX--$\beta$ diagram \citep{burgarella2005a,boquien2009b}. In effect when only GALEX bands are available in the UV, it is difficult to distinguish between the presence of a bump and an attenuation law with a shallower slope, as the bump is contained in the NUV band. Indeed, in some galaxies such as the Milky Way the presence of a bump around 220~nm has been widely observed, as well as at low \citep{burgarella2005a,conroy2010a,wild2011a} and high redshift \citep{noll2009b,buat2011b}. The strength of this bump strongly varies from galaxy to galaxy for reasons that are still subject of an intense debate in the literature. In addition to the effect of the metallicity, recent results suggest that the inclination plays a major role, with edge--on galaxies exhibiting a larger bump than face--on galaxies \citep{conroy2010a,wild2011a}.  Even though CIGALE can be used to detect the presence of a bump in the attenuation law \citep{buat2011b}, such detailed constraints require a fine sampling of the SED in rest--frame UV to quantify its amplitude accurately. In the context of the present study, the amplitude of the bump is expected to be weak as the selected galaxies are reasonably face--on \citep{conroy2010a,wild2011a} and fairly constant as the galaxies have similar metallicities ($<12+\log O/H>=8.66\pm0.11$, Hughes et al. 2012, in preparation), close to that of the solar neighbourhood \citep{rudolph2006a}. While the presence of a bump mainly affects the NUV band, the slope of the attenuation has a broader effect from the FUV to the optical. To model a variation of the slope, CIGALE uses the starburst attenuation law as a baseline, which is multiplied by a factor $\left(\lambda/\lambda_0\right)^\delta$, with $\lambda$ the wavelength, $\lambda_0$ the normalisation wavelength, and $\delta$ the slope modifying parameter. Therefore, $\delta=0$ corresponds to a starburst attenuation curve, $\delta>0$ to a shallower one, and $\delta<0$ to a steeper one, closer to the extinction law observed in the Magellanic Clouds for instance. To test whether the presence of a bump and a varying attenuation law slope have an effect on the results, we have examined the quality of the fit with several CIGALE runs 1) adding a bump with an amplitude similar to that of the LMC2 supershell \citep{gordon2003a} while leaving $\delta$ free, and 2) setting the slope of the attenuation curve to a fixed value. Leaving the slope parameter free brings substantial improvements to the quality of the fit both globally (measured with the mean $\chi^2$) and in the UV (measured with the mean difference between the modelled and the observed FUV-NUV colour). Conversely forcing the presence of a bump slightly degrades the quality of the fits. Given the poor constraints, and the expectation that the bump should be weak, we have chosen not to consider the presence of a bump but we have left the slope as a free parameter ranging from $\delta=-0.4$ to $\delta=0$, with steps of 0.1, which reproduces the observations. We have also fixed the attenuation fraction, $f_{att}$, between the old and young stellar populations to $f_{att}=0.5$ which is commonly observed. A test run has been performed with $f_{att}=0.75$, showing there is little impact on the results. The attenuation of the young population in the V band ranges from 0.05~mag to 1.65~mag with steps of 0.15~mag.

The emission in the IR is handled by CIGALE using different sets of models or templates such as \cite{dale2002a} or \cite{chary2001a} for instance. For this paper we mainly require good estimates of the dust luminosity. As we are not aiming at studying the detail of the properties of the FIR emission, the choice of templates has little consequence on the results. Therefore we have chosen to use the \cite{dale2002a} set of templates that has been extensively tested with CIGALE. These templates are fine--tuned to describe the properties of local, normal galaxies such as the ones selected in this sample. They are parametrised by the exponent $\alpha$ as defined in \cite{dale2002a}: $dM_d\left(U\right)\propto U^{-\alpha}dU$, with $M_d\left(U\right)$ the dust mass heated by a radiation field $U$. We sample this parameter between $\alpha=0.5$ and $\alpha=4.0$ with steps of 0.5, which covers the parameter space sampled by star--forming galaxies including extreme cases.

\begin{figure}[!htbp]
\centering
\includegraphics[width=\columnwidth]{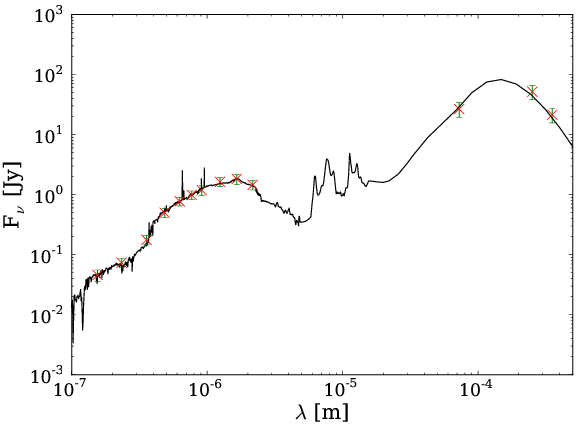}
\caption{Typical best fit SED by CIGALE from the FUV to the SPIRE 350~$\mu$m bands. This SED corresponds to one pixel in NGC~4254. The observed fluxes are represented by the red crosses with green 3--$\sigma$ error bars. The value of $\chi^2$ is 0.76. The parameters are derived taking into account the probability distribution function yielded by a set of reasonably good fits.}
\label{fig:fit-example}
\end{figure}

An example of a typical best--fit model found by CIGALE is presented in Fig.~\ref{fig:fit-example}.

\subsection{Accuracy on the output parameters\label{ssec:accuracy}}

The CIGALE code has been extensively tested by \cite{noll2009a,buat2011a,giovannoli2011a} on entire galaxies at low and high redshift. However, individual regions in galaxies are more likely to undergo local effects which can be explored with CIGALE.
To ensure that the fitting procedure is reliable for individual regions in galaxies we require 1) that the observed SED can be reproduced by the models, and 2) that the intrinsic parameters are accurately estimated. To ascertain whether these requirements are met we apply the methods developed in \cite{giovannoli2011a} and \cite{buat2011a}.

\begin{figure*}[!htbp]
\centering
\includegraphics[width=\columnwidth]{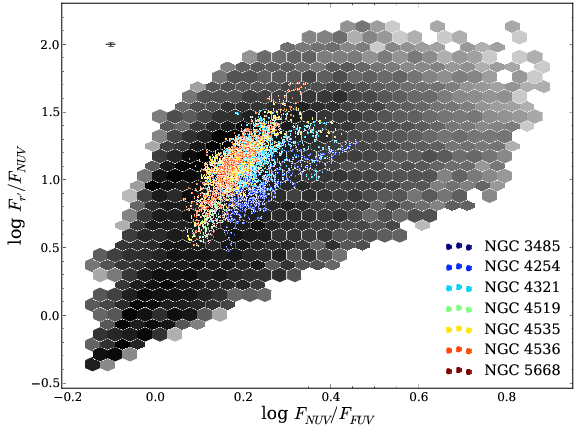}
\includegraphics[width=\columnwidth]{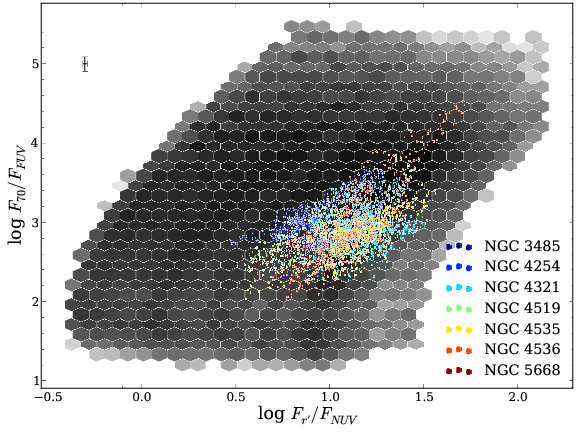}
\caption{Comparison between the colours of the models (grey hexagons) and the observed colours (circles). For better visibility, the grey shade is proportional to the log of the number of models within a given bin. Black represents a high density of models whereas white indicates that there is no model in this bin. The colour of the circles identifies the galaxy they belong to. The left plot compares the NUV$-r'$ versus the FUV$-$NUV colours whereas the right plots compares the FUV$-$MIPS 70 versus the NUV$-r'$ colours. The median 3--$\sigma$ error bars are displayed at the top--left corner of each plot.}
\label{fig:priors}
\end{figure*}

First, to test whether the models can reproduce the observations we compare the range of colours covered by the models to the colours of the observations. We concentrate on colours that constrain the star formation, attenuation, and properties of the IR emission. We compare the FUV-NUV, NUV-$r'$, and $r'$-MIPS 70 colours in Fig.~\ref{fig:priors}.
The set of models chosen covers a larger range than the observations which indicates that the observations can be reproduced by the models.

\begin{figure*}[!htbp]
\centering
\includegraphics[width=0.66\columnwidth]{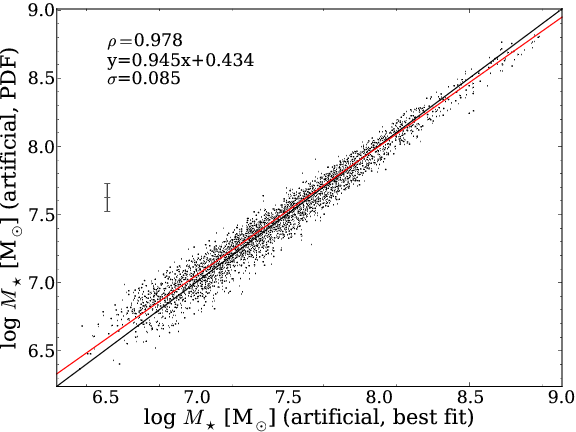}
\includegraphics[width=0.66\columnwidth]{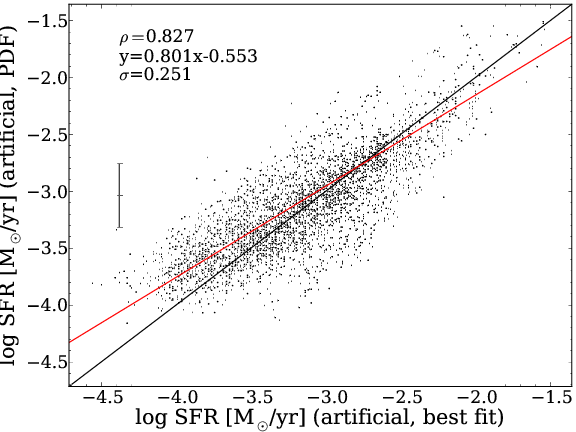}
\includegraphics[width=0.66\columnwidth]{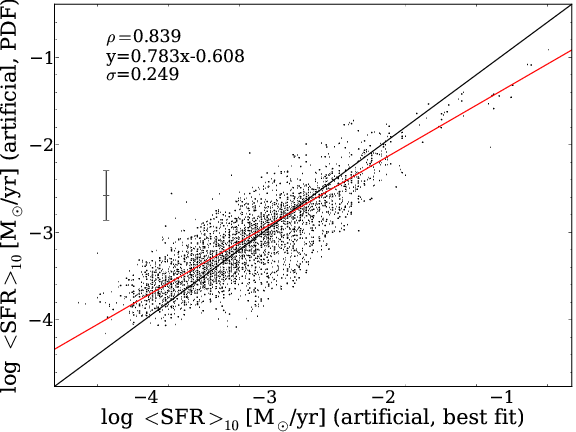}
\includegraphics[width=0.66\columnwidth]{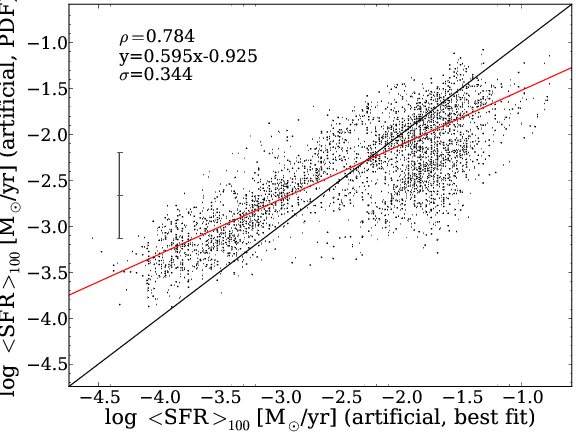}
\includegraphics[width=0.66\columnwidth]{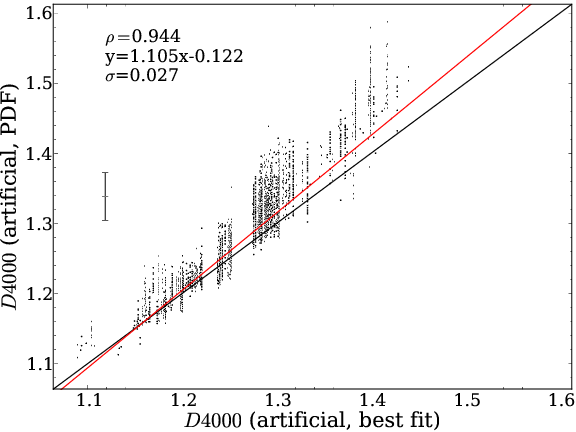}
\includegraphics[width=0.66\columnwidth]{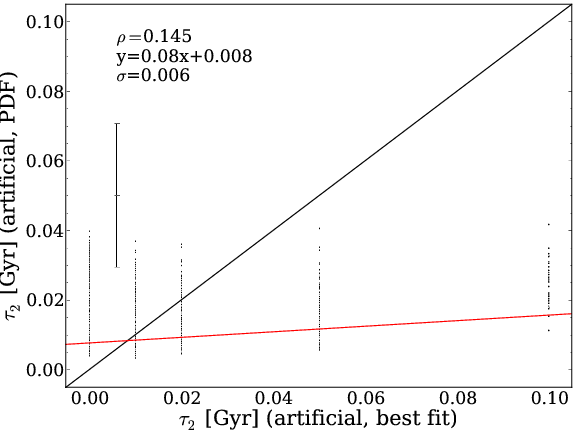}
\includegraphics[width=0.66\columnwidth]{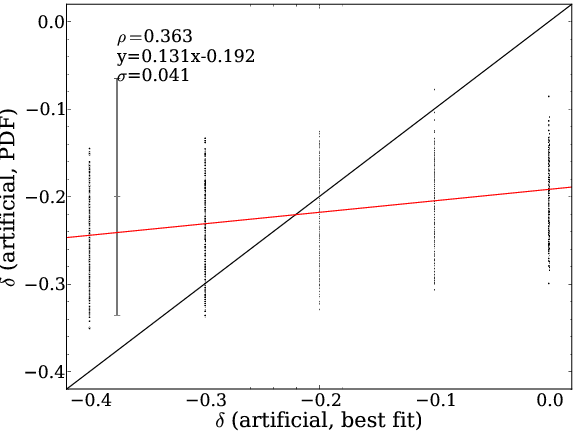}
\includegraphics[width=0.66\columnwidth]{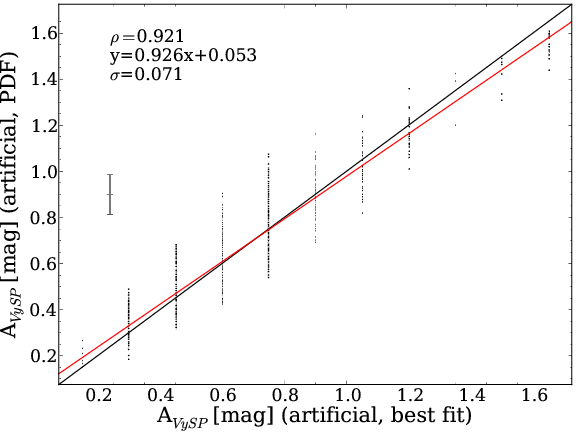}
\includegraphics[width=0.66\columnwidth]{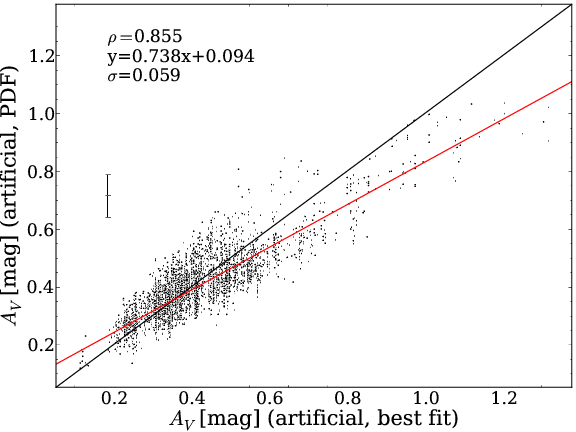}
\includegraphics[width=0.66\columnwidth]{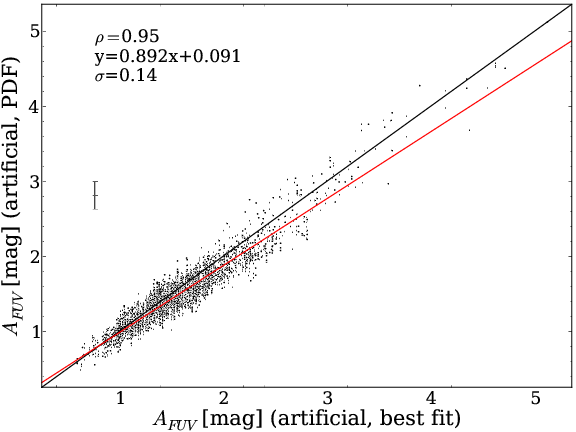}
\includegraphics[width=0.66\columnwidth]{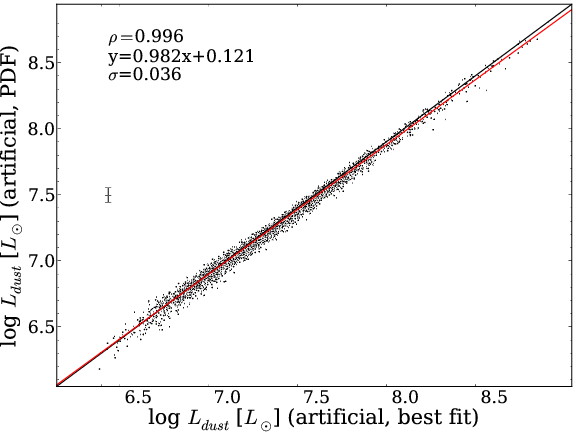}
\includegraphics[width=0.66\columnwidth]{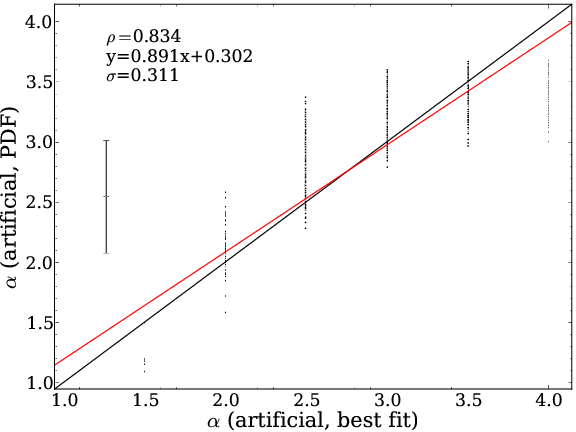}
\caption{Output parameters of the artificial catalogue for the best fit (x--axis) and the probability distribution function (PDF) analysis (y--axis) after a random error has been injected into the model SED. From the top--left corner to the bottom right corner the output parameters presented here are: log $M_\star$, log SFR, log $<$SFR$>_{10}$ (averaged over 10~Myr), log $<$SFR$>_{100}$ (averaged over 100~Myr), $D4000$, log $\tau_2$, $\delta$, $A_{VySP}$, $A_V$, $A_{FUV}$, log $L_{dust}$, and $\alpha$. The Pearson correlation coefficient $\rho$ is indicated in the top--left corner of each plot. For each parameter we have also computed the best linear fit which is shown in red along with the equation in the top--left corner where the standard deviation $\sigma$ around this best fit is also indicated. These fits take into account the uncertainties computed by CIGALE on each data point. The median error bar computed by CIGALE is shown on the left side of each plot.}
\label{fig:models-test}
\end{figure*}

Another important test is to check whether CIGALE reliably estimates the output parameters. This requires an a priori knowledge of the intrinsic characteristics of the galaxies. One method is to create an artificial catalogue of galaxies, following \cite{giovannoli2011a}. To do so we compute the best fit for each pixel in each galaxy. This first step yields a set of artificial SED including their intrinsic parameters given by CIGALE. To take into account the uncertainties on the observations and on the models, for each band we add a random uncertainty drawn from a gaussian distribution with $\sigma$ taken as the observed uncertainty computed on the original images. The latter is taken as the uncertainty on the artificial fluxes. We then perform a new run of CIGALE on the artificial catalogue, using the same input parameters as previously. In Fig.~\ref{fig:models-test} we compare the parameters of the artificial catalogue computed from the probability distribution function to the ones determined from the best fit along with the Pearson correlation coefficient $\rho$. The parameters are:
\begin{itemize}
 \item The stellar mass (log $M_\star$).
 \item The instantaneous SFR (log SFR), the SFR averaged over 10~Myr (log $<$SFR$>_{10}$) and 100 Myr (log $<$SFR$>_{100}$).
 \item The age--sensitive $D4000$ index defined as the ratio between the average flux density in the 400-410~nm  range and that in the 385--395~nm range.
 \item The e--folding time of the youngest star--forming episode ($\tau_2$).
 \item The attenuation curve slope modifying parameter ($\delta$).
 \item The V--band attenuation of the young ($A_{V_{ySP}}$) and total stellar populations ($A_{V}$), and the FUV attenuation ($A_{FUV}$).
 \item The dust luminosity (log $L_{dust}$).
 \item The $\alpha$ parameter as defined in \cite{dale2002a}, tracing the dust temperature.
\end{itemize}

The comparison between the results from the fit and the exact value of each parameter shows that CIGALE yields excellent results for most parameters with $\rho>0.8$ for log $M_\star$, log SFR, log $<$SFR$>_{10}$,  $D4000$, $A_V$, $A_{V,ySP}$, $A_{FUV}$, and log $L_{dust}$. For 5 of them $\rho>0.9$. Conversely, $\delta$ is poorly constrained ($\rho$=0.363) and log $\tau_2$ not at all ($\rho$=0.145). CIGALE yields nearly perfect estimates of the dust luminosity. This is due to the fact that having observations below and beyond the peak of the dust emission constrains the location and the amplitude of this peak which emits the bulk of the dust luminosity. Not having observations at shorter wavelengths has therefore only a minimal effect.

\section{Results and discussion\label{sec:discussion}}

\subsection{IRX--\texorpdfstring{$\beta$}{beta} diagrams and influence of the physical parameters}

As mentioned in Sec.~\ref{sec:introduction}, the IRX--$\beta$ diagram is one of the key tools to correct star--forming galaxies for the attenuation, linking $\beta$ to $IRX\equiv\log\left(L_{dust}/L_{FUV}\right)$. Both terms are intimately related to the attenuation.

First of all, the difference between the observed UV slope $\beta$ and the intrinsic UV slope in the absence of dust $\beta_0$, that is the reddening of the UV slope by the presence of dust, can be directly connected to the effective attenuation:
\begin{equation}
A_{FUV}=a_\beta\left(\beta-\beta_0\right).\label{eqn:Afuv-beta}
\end{equation}
This relation naturally separates the effect of the SFH which entirely and exclusively determines $\beta_0$, from the effect of the shape of the attenuation curve which entirely and exclusively determines $a_\beta$: 
\begin{equation}
a_\beta=\frac{k_{FUV}}{\left(k_{FUV}-k_{NUV}\right)\times2.5\log \lambda_{FUV}/\lambda_{NUV}},\label{eqn:a-beta}
\end{equation}
with $k$ the attenuation curve. In other words, $a_\beta$ represents the sensitivity of the FUV attenuation to the reddening by the dust.

At the same time IRX quantifies the relative fraction of the UV radiation reprocessed by dust, which permits us, under some approximation, to estimate the attenuation: \begin{equation}
A_{FUV}=2.5\log\left(1+a_{IRX}10^{IRX}\right),\label{eqn:Afuv-IRX}
\end{equation}
with $a_{IRX}$ a constant that depends on the relative amount of emission in the FUV band compared to attenuation sensitive bands \citep{meurer1999a}. Thus, combining equations \ref{eqn:Afuv-beta} and \ref{eqn:Afuv-IRX}, $\beta$ and IRX are linked through the following equation:
\begin{equation}
IRX=\log\left[\left(10^{0.4a_\beta\left(\beta-\beta_0\right)}-1\right)/a_{IRX}\right]. \label{eq:IRX}
\end{equation}
A full derivation of equation \ref{eq:IRX} is presented in \cite{hao2011a}.

\begin{figure*}[!htbp]
\centering
\includegraphics[width=0.66\columnwidth]{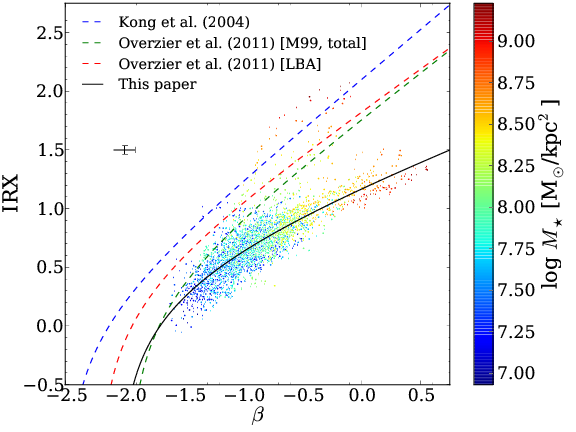}
\includegraphics[width=0.66\columnwidth]{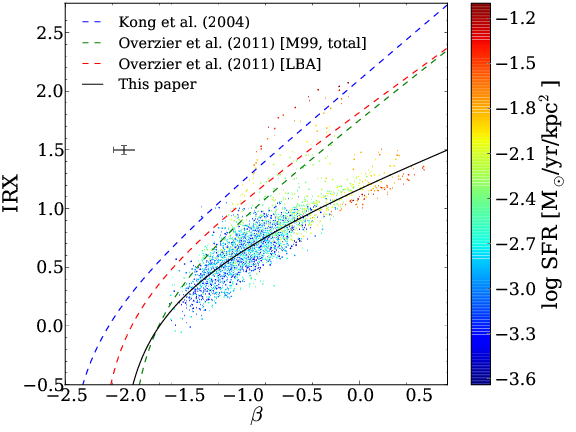}
\includegraphics[width=0.66\columnwidth]{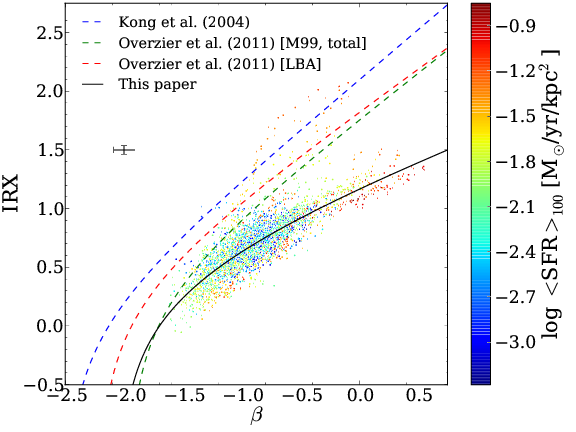}
\includegraphics[width=0.66\columnwidth]{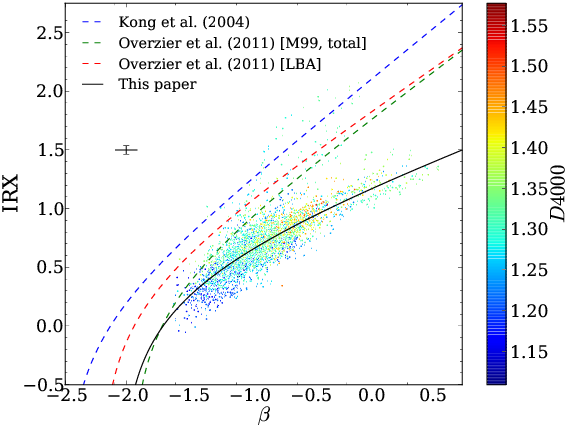}
\includegraphics[width=0.66\columnwidth]{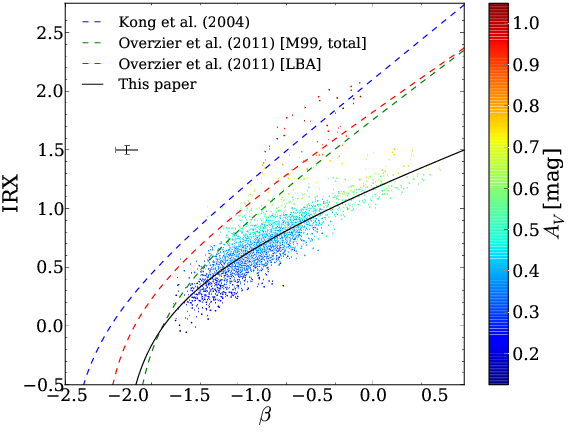}
\includegraphics[width=0.66\columnwidth]{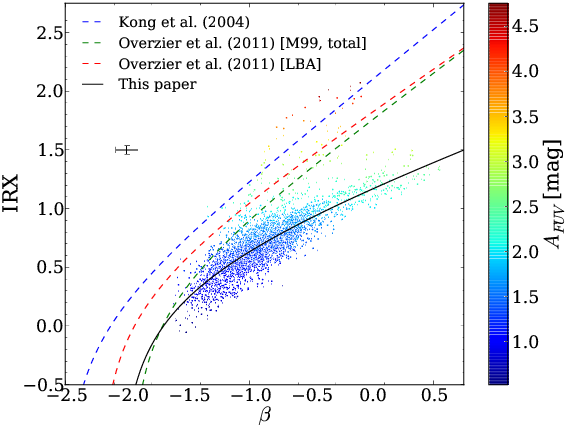}
\includegraphics[width=0.66\columnwidth]{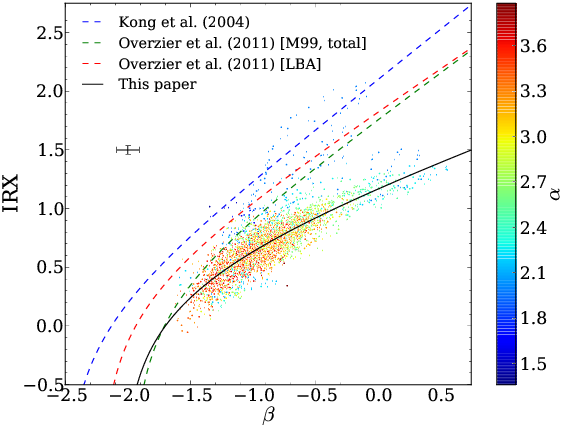}
\includegraphics[width=0.66\columnwidth]{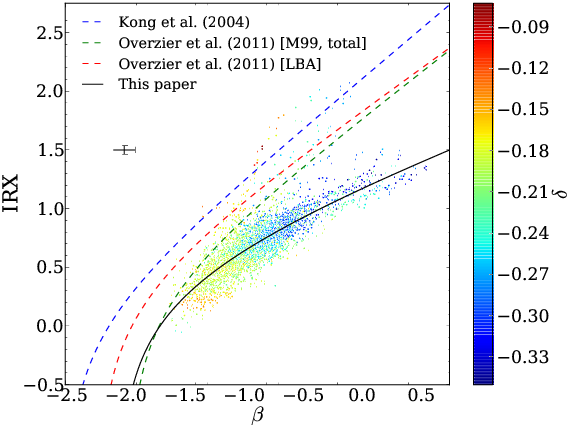}
\includegraphics[width=0.66\columnwidth]{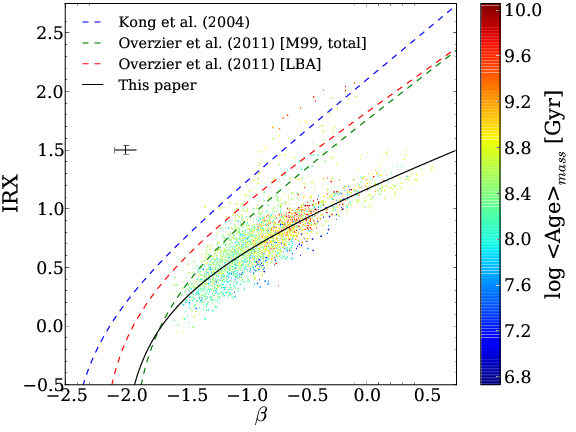}
\caption{Ratio between the IR (provided by CIGALE) and observed FUV luminosities, IRX (y--axis) versus the UV slope $\beta$ (x--axis). The colours of the individual points represent the value of the parameter indicated on the right of the colourbar. From the top--left corner to the bottom--right corner the parameters presented here are: log $M_\star$, log SFR, log $<$SFR$>_{100}$ (averaged over 100~Myr), $D4000$ index, $A_V$, $A_{FUV}$, $\alpha$, $\delta$, and log $<$Age$>_{mass}$. Blue points indicate a low value whereas red points indicate a high value. The dashed lines represent the IRX--$\beta$ relations of \cite{kong2004a} in blue, and the \cite{meurer1999a} and Lyman--break analogue relations (green and red) that have been derived by \cite{overzier2011a}. The median 1--$\sigma$ uncertainty is displayed on the left side of each plot. Finally, the solid black line represents the best fit for the entire sample, minus the discarded data points: $IRX=\log\left[\left(10^{0.990\left(\beta+2.046\right)}-1\right)/0.373\right]$.}
\label{fig:irx-beta}
\end{figure*}

The observed slope $\beta$ is not directly accessible to us as only broadband UV observations are available. For a given SED, the value of $\beta$ can vary depending on the available bands, due to colour effects generated by the shapes of the filters. We compute $\beta$ from the GALEX FUV and NUV bands using the following relation:
\begin{equation}
\beta=\frac{\log\left(F_{NUV}/F_{FUV}\right)}{\log\left(\lambda_{FUV}/\lambda_{NUV}\right)}-2,
\end{equation}
with $F$ the flux density and $\lambda$ the wavelength. It can also be expressed in terms of magnitudes:
\begin{equation}
 \beta=\frac{FUV-NUV}{2.5\log\left(\lambda_{FUV}/\lambda_{NUV}\right)}-2.
\end{equation}
To compute $IRX\equiv\log\left(L_{dust}/L_{FUV}\right)$ we use the dust luminosity provided by CIGALE, which directly comes from the energy balance between dust absorption and emission. We present some IRX--$\beta$ diagrams in Fig.~\ref{fig:irx-beta} showing the data for all the galaxies in order to examine the influence of various physical parameters estimated by CIGALE on the IRX--$\beta$ relation.

In Fig.~\ref{fig:irx-beta}, we see that the points from the selected HRS galaxies lie well below the IRX--$\beta$ starburst relations derived by \cite{overzier2011a}, which is in accordance to what is expected for non starbursting galaxies \citep{bell2002a,buat2002a,kong2004a,gordon2004a,seibert2005a,calzetti2005a,boissier2007a,dale2007a,johnson2007a,panuzzo2007a,cortese2008a,munoz2009a,boquien2009b}. There are a number of points that present a significantly higher IRX at a given $\beta$ in comparison to the envelope described by the points. These points are roughly compatible with the relations defined for starburst galaxies. A close inspection reveals they belong to NGC~4536 which is undergoing a nuclear starburst. This shows that several different regimes in terms of IRX--$\beta$ can coexist within a single galaxy. We also see that the location in the IRX--$\beta$ diagram is strongly linked to some of the parameters. For instance, the stellar mass shows a clear gradient along the envelope. 

\begin{figure}[!htbp]
\centering
\includegraphics[width=\columnwidth]{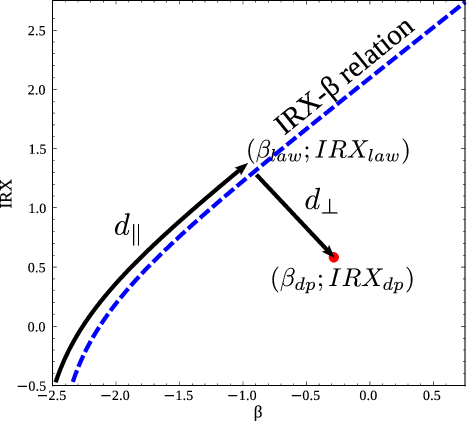}
\caption{Diagram showing how $d_\perp$ and $d_\parallel$ are computed between a data point of coordinates ($\beta_{dp};IRX_{dp}$) and a given IRX--$\beta$ relation.}
\label{fig:irx-beta-distance}
\end{figure}

To understand the different trends with the parameters and the deviation from the IRX--$\beta$ starburst relation from the literature, we have defined 2 quantities: the perpendicular and parallel distances $d_\perp$ and $d_\parallel$ as shown in Fig.~\ref{fig:irx-beta-distance}.
The perpendicular distance quantifies the deviation away from a given IRX--$\beta$ relation whereas the parallel distance quantifies gradients along the relation.
By convention, data points that are located above (respectively under) the curve have $d_\perp<0$ (resp. $d_\perp>0$).
To compute $d_\parallel$ we choose the origin of the IRX--$\beta$ curve to be set at IRX=$-0.5$. The choice of the origin has no influence on the results. As a reference curve we use the IRX--$\beta$ relation of \cite{kong2004a}: $IRX=\log\left(10^{2.1+0.85\beta}-0.95\right)$\footnote{Note that a relation of the form $IRX=\log\left(10^{a\beta+b}-c\right)$ is easily converted to a form similar to equation \ref{eq:IRX}, with $a_{IRX}=1/c$, $\beta_0=\left(\log\left(c\right)-b\right)/a$, and $a_\beta=2.5\times a$.}. The computed values of the Spearman correlation coefficient of $d_\perp$ and $d_\parallel$ versus various output parameters are provided in Table~\ref{tab:distances} along with the birthrate parameter $b$, the ratio of the current to the average SFR over the lifetime of the galaxy, which is also equivalent to the specific SFR. The data points affected by the starbursting region in NGC~4536 have been discarded by only selecting points with $IRX<0.5\beta+1.5$. The plots are provided in the appendix, in Fig.~\ref{fig:dist-perp} and \ref{fig:dist-para}

\begin{table}
\caption{Spearman correlation coefficients for $d_\perp$ and $d_\parallel$ versus various output parameters.}
\label{tab:distances}
\centering
\begin{tabular}{c c c}
 \hline\hline
Parameter & $\rho_{d_\perp}$ & $\rho_{d_\parallel}$\\
\hline
$A_{FUV}$          &$ 0.07$&$ 0.86$\\
$A_{NUV}$          &$ 0.06$&$ 0.85$\\
$A_{V_{ySP}}$      &$-0.02$&$ 0.82$\\
$A_{V}$            &$ 0.03$&$ 0.76$\\
$<age>_{mass}$     &$ 0.05$&$ 0.35$\\
$\alpha$           &$-0.19$&$-0.35$\\
$\delta$           &$-0.29$&$-0.62$\\
$D4000$            &$ 0.04$&$ 0.36$\\
$\log SFR$         &$ 0.16$&$ 0.38$\\
$\log <SFR>_{10}$  &$ 0.15$&$ 0.36$\\
$\log <SFR>_{100}$ &$ 0.17$&$ 0.09$\\
$\log L_{dust}$    &$ 0.27$&$ 0.51$\\
$\log L_{bol}$     &$ 0.33$&$ 0.51$\\
$\log M_\star$     &$ 0.29$&$ 0.60$\\\hline
$\log b$           &$-0.25$&$-0.34$\\
$\log <b>_{10}$    &$-0.24$&$-0.33$\\
$\log <b>_{100}$   &$-0.11$&$-0.39$\\
\hline
\end{tabular}
\end{table}

The various parameters considered are mostly uncorrelated with $d_\perp$. However, we see some weak structures at higher values of $d_\perp$ with some data points that have a higher attenuation, SFR, bolometric, and dust luminosities  Conversely $d_\parallel$ presents a convincing correlation with the majority of parameters.
There is no trend with the birthrate parameter, either instantaneous or averaged over 10~Myr or 100~Myr. Physically, this means that none of these parameters seems to be directly responsible for the deviation from the starburst IRX--$\beta$ relation, at least on a sub--galactic scale.

\subsection{Relation between IRX and the attenuation}

IRX and attenuation are closely linked with quantitative relations calibrated on entire galaxies widespread in the literature \citep{kong2004a,buat2005a,burgarella2005a,cortese2008a,hao2011a}. Whether these relations are also valid for individual regions in galaxies is still an open question. CIGALE provides us with several measures of the attenuation: $A_{FUV}$, $A_{NUV}$, $A_{V_{ySP}}$, and $A_V$. It is difficult to convert analytically from one attenuation to the other as they depend not only on the underlying attenuation law, but also on the relative luminosity of the two populations in each band. As the relation provided in equation \ref{eqn:Afuv-IRX} is not valid for longer wavelengths, following \cite{cortese2008a,buat2011a}, we have derived the relation between the attenuation and IRX using a fourth order polynomial, giving all relations the same analytic form.

\begin{eqnarray*}
A_{FUV}&=&0.0806x^4-0.1922x^3+0.6009x^2+0.8071x\\&&+0.6621\ (\sigma=0.109),\\
A_{NUV}&=&0.1086x^4-0.2986x^3-0.5621x^2+0.5910x\\&&+0.4814\ (\sigma=0.084),\\
A_{V_{ySP}}&=&-0.0382x^4+0.2322x^3-0.1732x^2+0.5341x\\&&+0.1901\ (\sigma=0.043),\\
A_{V}&=&0.1388x^4-0.3451x^3+0.3130x^2+0.2302x\\&&+0.1759\ (\sigma=0.046),
\end{eqnarray*}
where $x\equiv\mathrm{IRX}$ and $\sigma$ is the standard deviation around the best fit. Discrepant points in NGC~4536 have been discarded.

\begin{figure}[!htbp]
\centering
\includegraphics[width=\columnwidth]{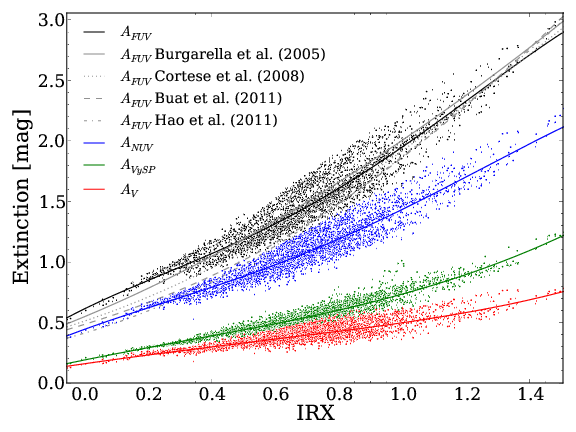}
\caption{$A_{FUV}$ (black points), $A_{NUV}$ (blue points), $A_{V_{ySP}}$ (green points), and $A_V$ (red points) versus IRX. The solid lines represent the best fit with a fourth order polynomial for each attenuation measurement. The gray lines represent estimates of $A_{FUV}$ published in the literature. Discrepant points in NGC~4636 have been discarded.}
\label{fig:extinction-irx}
\end{figure}

In Fig.~\ref{fig:extinction-irx} we plot the relations aforementioned between IRX and the attenuation measurements.

The relation derived for $A_{FUV}$ gives results close to the ones found in the literature for entire galaxies. Computing $\Delta A_{FUV}$ as the difference between the relation derived in this paper and the relation published in the literature, we find:

\begin{itemize}
 \item $\Delta A_{FUV}=-0.028\pm0.032$ \citep{burgarella2005a},
 \item $\Delta A_{FUV}=0.068\pm0.031$ \citep{cortese2008a},
 \item $\Delta A_{FUV}=0.138\pm0.039$ \citep{buat2011a},
 \item $\Delta A_{FUV}=0.156\pm0.056$ \citep{hao2011a},
\end{itemize}
The fact that we find relations close to those derived using entire galaxies shows that the effect of radiation transfer on results is at most minimal.

\subsection{Relation between \texorpdfstring{$\beta$}{beta} and the attenuation\label{sssec:beta-AFUV}}

\begin{figure}[!htbp]
\centering
\includegraphics[width=\columnwidth]{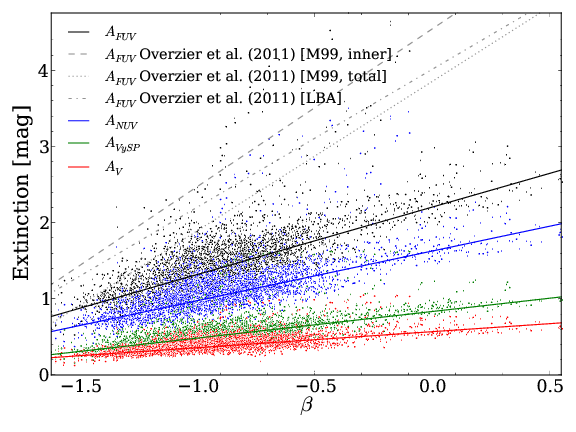}
\caption{$A_{FUV}$ (black points), $A_{NUV}$ (blue points), $A_{V_{ySP}}$ (green points), and $A_V$ (red points) versus $\beta$. The solid lines represent the best linear fit for each attenuation measurement after the discrepant points from NGC 4536 have been discarded. The gray lines represent estimates of $A_{FUV}$ published in the literature.}
\label{fig:extinction-beta}
\end{figure}

As mentioned earlier, assuming the unextinguished colour of the UV dominating population $\beta_0$ is constant, the attenuation is related to the reddening of UV colours by equation \ref{eqn:Afuv-beta}: A$_{FUV}=a_\beta\left(\beta-\beta_0\right)$. In Fig.~\ref{fig:extinction-beta}, we plot the relation between $\beta$ and the attenuation.

We obtain the following relations assuming $a_\beta$ and $\beta_0$ constant:
\begin{eqnarray*}
A_{FUV}&=&0.870\left(\beta+2.586\right)\ (\sigma=0.216),\\
A_{NUV}&=&0.633\left(\beta+2.616\right)\ (\sigma=0.162),\\
A_{V_{ySP}}&=&0.337\left(\beta+2.524\right)\ (\sigma=0.093),\\
A_{V}&=&0.203\left(\beta+2.868\right)\ (\sigma=0.067),
\end{eqnarray*}
where $\sigma$ is the standard deviation around the best fit. These relations do not take into account the discrepant points from NGC~4536.

Compared to the relations for starburst galaxies presented in \cite{overzier2011a}, the relations derived from the selected sample 1) yield a systematically lower attenuation for a given $\beta$, and 2) have a shallower slope. The first point is not surprising as it is a well known fact that star--forming galaxies lie below the starburst IRX--$\beta$ relation \citep{kong2004a}, which is also the case for entire HRS star--forming galaxies as we will see in Sec.~\ref{ssec:galaxies}. The parameter $a_\beta$ in equation~\ref{eqn:Afuv-beta} is directly linked to the shape of the attenuation law (equation~\ref{eqn:a-beta}). The value $a_\beta=0.870$ corresponds to a particularly extreme attenuation law. A starburst attenuation law ($\delta=0$) yields a rather grey slope of $\sim2.3$, while $\delta\sim-0.2$ (resembling the LMC2/supershell extinction curve, excluding the presence of a bump) yields a slope of $\sim1.7$, and $\delta\sim-0.5$ (resembling an SMC--like extinction curve, excluding the presence of a bump) yields a slope of $\sim1.3$. This shows that assuming that $\beta_0$ is a constant, the shallow slope of the $A_{FUV}$--$\beta$ relation would yield an unphysically steep attenuation law. At the same time we notice that there is a considerable scatter around the fit. These observations suggest that $\beta_0$ might vary significantly across the sample depending of the actual SFH of each data point. We will examine this possibility in detail in Sec.~\ref{ssec:deviation}.

The low, unphysical value of $a_\beta$ has a large impact when using starburst IRX--$\beta$ relations to correct for the attenuation. Comparing the ``M33 inner'' relation obtained by \cite{overzier2011a}, the $A_{FUV}$ attenuation would be overestimated by 0.6~mag for $\beta=-1.5$, 1.2~mag for $\beta=-1$, 1.7~mag for $\beta=-0.5$, and 2.3~mag for $\beta=0$, yielding errors on the SFR up to nearly an order of magnitude, which would overestimate the contribution of normal star--forming galaxies to the cosmic star formation.

\subsection{Impact of the variation of \texorpdfstring{$\beta_0$}{the intrinsic UV slope}\label{ssec:deviation}}

\subsubsection{Effect of the star formation history}

\begin{figure*}[!htbp]
\centering
\includegraphics[width=\columnwidth]{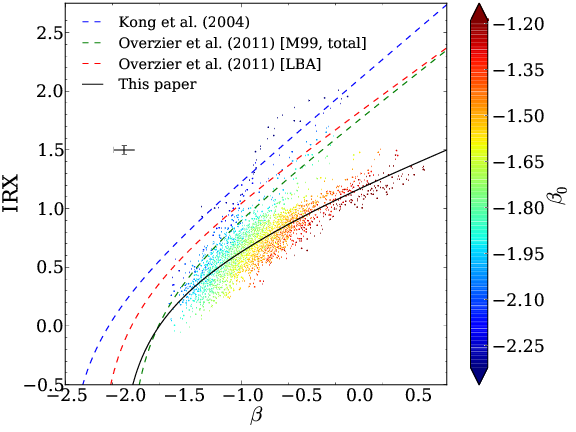}
\includegraphics[width=\columnwidth]{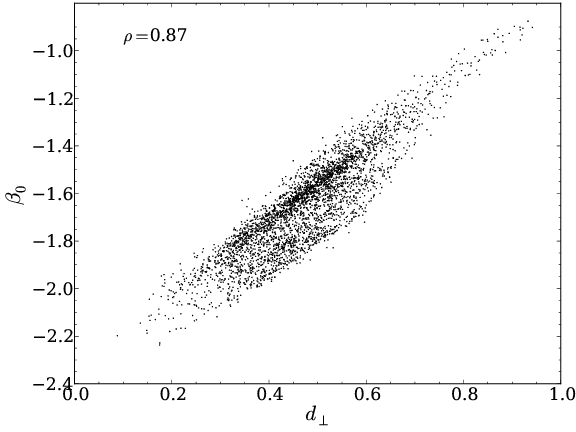}
\caption{Left: Same as Fig.~\ref{fig:irx-beta} with the colour of each data point corresponding to the intrinsic UV slope $\beta_0$, ranging from $-2.32$ to $-1.19$. Data points lower than $-2.32$ (resp. higher than $-1.19$) are shown in dark blue (resp. dark red) allowing a larger dynamic range for the bulk of the data points. Right: Intrinsic UV slope $\beta_0$ versus the perpendicular distance $d_\perp$. The Spearman correlation coefficient is $\rho=0.87$.}
\label{fig:beta0-effect}
\end{figure*}

In galaxies forming stars at a lower rate, the disk--averaged SFR can be low enough so that the intrinsic shape of the SED in the UV and optical bands differs significantly not only from that of a starburst galaxy but also from object to object. Such an effect was explored by \cite{kong2004a} in terms of the birthrate parameter $b$. Previous work \citep{cortese2006b,johnson2007a,cortese2008a,boquien2009b} also found a possible effect from the age of the populations. However the values of $b$, the $U-B$ colour, the $D4000$ index, or the H$\alpha$ equivalent width that have been used in the literature to quantify the age do not provide an estimate of the shape of the SED in the UV domain in a direct way. Indeed, they have significantly different timescale sensitivities and are affected by attenuation at different levels. Through careful modelling of the SED, CIGALE provides us with direct estimates of the reddening underwent by the intrinsic UV slope $\beta_0$ which allows us to correct $\beta$ for the attenuation in order to retrieve $\beta_0$. In Fig.~\ref{fig:beta0-effect} (left) we present the IRX--$\beta$ diagram with each data point colour coded according to $\beta_0$.
We observe that there is a clear gradient of $\beta_0$ according to the perpendicular distance from the starburst law of \cite{kong2004a} as confirmed in Fig.~\ref{fig:beta0-effect} (right). This strongly hints that intrinsic differences in the UV slope play a major role to explain the deviation from the starburst relation. At the same time there is also a non negligible scatter around the relation which suggests that other parameters such as the shape of the attenuation law also play a role.

The shape of the attenuation curve also constrains the $\beta$--$A_{FUV}$ relation and the location of the data points in the IRX--$\beta$ diagram \citep{hao2011a}. Indeed, a steeper slope will increase the reddening of the UV slope for a given quantity of dust attenuation while absorbing a larger fraction of the emission at shorter wavelength. Conversely, the presence of a bump in the NUV band will reduce the FUV-NUV differential attenuation while only slightly increasing the fraction of reprocessed UV light into the FIR. As we have shown not all data points have the same $\beta_0$, therefore a simple fit of the data points does not yield any direct information on the attenuation curve. At the same time, in Fig~\ref{fig:irx-beta} we see that data points that have a high $\beta$ also tend to have a low $\delta$, that is a steeper attenuation curve slope. There is a visible correlation between $\delta$ and $d_\parallel$ ($\rho=-0.62$). The variation of $\beta_0$ makes it difficult to evaluate changes in the attenuation law across the entire sample. Therefore, to test how variations in the attenuation law affect the estimate of $A_{FUV}$, we have divided the sample in 10 bins of $\beta_0$ and determined the $\beta$--$A_{FUV}$ relation in each of these bins such that $A_{FUV}=a_\beta\left(\beta-\beta_0'\right)$, with $a_0$ and $\beta_0'$ constants derived from the fit of this relation. For a fixed $\beta_0$, the relation $A_{FUV}$--$\beta$ only depends on the shape of the attenuation curve as we saw in equation~\ref{eqn:a-beta}. The narrower range of $\beta_0$ limits the effects of the SFH and allows us to examine if and how $\beta_0$ and $\delta$ are linked.

The parameters of this relation are summarised in Table~\ref{tab:best-fit-bins}.
\begin{table}
\caption{Best fit $A_{FUV}=a_\beta\left(\beta-\beta_0'\right)$ in different bins of $\beta_0$.}
\label{tab:best-fit-bins}
\centering
\begin{tabular}{c c c}
 \hline\hline
bin & $a_\beta$ & $\beta_0'$\\
\hline
$\beta_0\le-2.32$      &$2.28\pm0.15$&$-2.28\pm0.10$\\
$-2.32<\beta_0\le-2.18$&$2.09\pm0.06$&$-2.16\pm0.03$\\
$-2.18<\beta_0\le-2.04$&$2.07\pm0.05$&$-2.02\pm0.02$\\
$-2.04<\beta_0\le-1.90$&$1.93\pm0.02$&$-1.94\pm0.01$\\
$-1.90<\beta_0\le-1.76$&$1.83\pm0.02$&$-1.85\pm0.01$\\
$-1.76<\beta_0\le-1.62$&$1.70\pm0.02$&$-1.77\pm0.01$\\
$-1.62<\beta_0\le-1.47$&$1.63\pm0.02$&$-1.68\pm0.01$\\
$-1.47<\beta_0\le-1.33$&$1.67\pm0.02$&$-1.52\pm0.01$\\
$-1.33<\beta_0\le-1.19$&$1.66\pm0.03$&$-1.40\pm0.03$\\
$-1.19<\beta_0$        &$1.48\pm0.05$&$-1.37\pm0.05$\\
\hline
\end{tabular}
\end{table}
We find that data points that have $\beta_0\lesssim-1.8$, that is the bluest UV slope, have values of $a_\beta$ that are similar to those of starburst galaxies as determined by \cite{overzier2011a} ($1.81\le a_\beta\le2.07$). Conversely, bins that have a higher $\beta_0$ tend to have a smaller $a_\beta$, which indicates that the effective attenuation law is steeper. The change from low to high $\beta_0$ corresponds to a steepening of the attenuation curve. It can be easily understood as a transition between a starburst law for strongly star--forming regions that resemble starburst galaxies to attenuation laws seen in star--forming galaxies. As the stellar populations age, the dust clouds are dispersed and coherent feedback ceases. The physical conditions and geometry required for a starburst attenuation law are no longer. Finally, we notice that the value of $\beta_0'$ determined by the fit, while close to the value of $\beta_0$ obtained from CIGALE shows some deviations. This is likely due to variations of the attenuation law within each bin.
 
These results confirm that the primary reason star–forming galaxies deviate from the starburst relation is due to differences in their intrinsic UV colour, hence a different $\beta_0$.
This is most likely due to differing SFH from one galaxy to another on the timescale the UV is sensitive to. At the same time there is a clear variation of the slope of the attenuation curve, going from a starburst--like curve for regions in galaxies with a low $\beta_0$, and steepening with increasing $\beta_0$.

\subsubsection{Impact of the weak constraint on \texorpdfstring{$\delta$}{delta}}

As we have seen in Sec.~\ref{ssec:accuracy}, the constraint on $\delta$ is weak. In addition, the dynamical range on $\delta$ is compressed from 0.5 to $\sim0.25$ in artificial catalogues as can be seen in Fig.~\ref{fig:models-test}. In turn, this compressed dynamical range exacerbates the relation between $\beta_0$ and $d_\perp$. To obtain strong constraints on $\delta$, a good sampling of the UV continuum with medium or narrow--band filters from the FUV to the U band are required \citep[][Buat et al. 2012, in prep.]{buat2011b}, which is difficult to obtain for nearby galaxies as the UV is not observable from the ground. To ensure that our results are not affected by the weak constraint on $\delta$, we have performed several test runs. First of all, we have fitted each data point setting $\delta=-0.25$, which is typical of the value obtained when $\delta$ is set free. Qualitatively we find that the results are nearly identical and the quality of the fits while not as good remains excellent, showing the major influence of $\beta_0$. Conversely, we set $\beta_0$ to the typical value of $-1.7$ by selecting a SFH leading to such a value. We also set $\delta$ free to vary in a wide range well beyond realistic values, such as $\delta<-0.5$. The best fits we obtain are considerably worse than previously, with a significant number of them failing catastrophically and yielding large errors in particular in the UV and in the IR, sometimes over 1 magnitude. This shows that intrinsically, a variation of the attenuation law is not sufficient to explain why normal star--forming galaxies are located under the starburst IRX--$\beta$ curve. Conversely, variations in the SFH which naturally yield variations in $\beta_0$ can be sufficient to explain the deviation. In addition, as we saw in Sec.~\ref{sssec:beta-AFUV}, the relation between $A_{FUV}$ and $\beta$ could only be explained by an extreme attenuation law that would be even steeper than the SMC extinction law. This is particularly unlikely as the metallicity of the sample is close to the solar one. However, a spread in $\beta_0$ would naturally make the $A_{FUV}$--$\beta$ relation more shallow without requiring a variation of the attenuation law.

This shows that the poor constraint on $\delta$ from CIGALE does not really affect the results and that the variations of $\beta_0$ are the driving reason why normal star--forming galaxies are located under the starburst IRX--$\beta$ relation and that variations of the attenuation law play only a secondary role.

\subsubsection{Relation between \texorpdfstring{$\beta_0$}{the intrinsic UV slope} and the distance from the centre}

\begin{figure}[!htbp]
\centering
\includegraphics[width=\columnwidth]{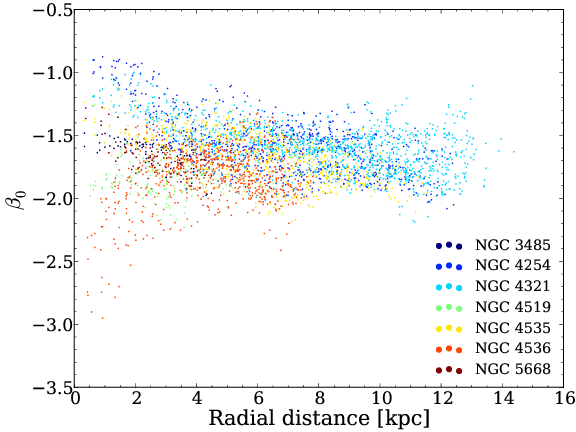}
\caption{Value of $\beta_0$ as a function of the inclination--corrected distance from the centre of the galaxy in kpc. The colour of the circles identifies the galaxy they belong to.}
\label{fig:correl-dist-beta0}
\end{figure}

If we examine the relation between $\beta_0$ and the distance from the centre of each galaxy, we notice that the outer regions tend to have a bluer $\beta_0$ than inner regions (Fig.~\ref{fig:correl-dist-beta0}).

Two effects can be at play here. First there can be metallicity gradients. The nucleus of a galaxy is generally more metal--rich than the outer parts. The presence and the strength of these gradients depends on its intrinsic parameters but also whether it is interacting or not as radial mixing flattens metallicity gradients \citep{barnes1992a,kewley2010a,rupke2010a,rupke2010b}. More metal--rich star--forming regions will have a redder $\beta_0$ because of the presence of numerous absorption lines in the photosphere of stars. The second effect that can generate a bluer $\beta_0$ in outer regions is simply due to the presence of numerous stars in the inner regions which can contaminate the UV colour making $\beta_0$ redder.

Disentangling these 2 effects is difficult and would require to have accurate metallicity maps for all the sample. To estimate the range of the effect, we have modelled a star--forming region with a metallicity of Z=0.02 and Z=0.001. Assuming a constant SFR over 10~Myr, we find that $\beta_0$ is $\sim0.3$~dex bluer in the low metallicity case. The difference is even smaller when considering a constant SFR over a longer timescale. Such a low metallicity is unlikely in the large spiral galaxies in the sample, especially that the emission in the FIR would become particularly faint due to the depletion of dust. This gives us an upper bound to the expected effect. We see in Fig.~\ref{fig:correl-dist-beta0} that for some of the galaxies, like NGC~4254, NGC~4321, or NGC~4535, the gradient of $\beta_0$ appears to be larger than what could be explained even by an extreme metallicity gradient. It shows that while a change of metallicity between inner and outer regions may play a role, it is not sufficient to explain the range of $\beta_0$.

\subsection{Impact of the choice of the pixel size\label{ssec:pixel-size}}

The choice of a 8\arcsec\ pixel size for a PSF of 24\arcsec\ could influence the results in case of improper convolution to the lower resolution. To test whether our results are affected by the pixel size, we have reprocessed the data choosing a pixel size of 24\arcsec, and performed a new analysis. It turns out that the influence is weak and that our results are not affected. The respective envelopes in the IRX--$\beta$ diagram and the gradients described by the two sets are similar. The correlation coefficients of the parameters with $d_\perp$ and $d_\parallel$ show a variation by typically no more than 0.1.

\subsection{Comparison with entire galaxies\label{ssec:galaxies}}

It is unclear whether the results we have obtained on individual data points within nearby galaxies also hold for normal star--forming galaxies. Indeed, if at a local level strong variations of $\beta_0$ are expected due to a quickly varying SFR, the intrinsic UV slope of entire galaxies is thought to undergo smaller variations as a much larger number of star--forming regions is averaged over the galactic disk compared to individual regions. This could for instance lead to a stronger role of a variation of the shape of the attenuation law.

Being a K--band selected, volume limited sample, the HRS contains a large number of normal spiral galaxies. Conversely, IR or UV selected samples can contain a significantly larger proportion of actively star--forming galaxies that are not necessarily representative of normal spiral galaxies. We have selected a subset of 63 HRS galaxies that have a Sa or later morphological type, excluding galaxies that are too inclined requiring $a/b<3$, with $a$ the major axis and $b$ the minor axis, to limit radiation transfer effects affecting the SED fitting. Compared to the study of individual regions, we forgo the SDSS $u'$ and $z'$ bands as well as the MIPS 70~$\mu$m one. Conversely we use ground--based U and V images that are deeper than their SDSS counterpart as well as IRAS 60~$\mu$m images as they offer a larger coverage of the HRS than MIPS 70~$\mu$m images. The range of CIGALE input parameters determined for individual star--forming regions is not necessarily adapted to entire galaxies. We have chosen to use the set of parameters determined by \cite{buat2011a}, with the difference that we allow $\tau_2$ to cover the same range of parameters as $t_2$. That is, we allow the latest star formation episode to be over. Examination of the priors similar to those presented in Fig.~\ref{fig:priors} shows that all observations of the subsample can be reproduced by the models.

\begin{figure}[!htbp]
\centering
\includegraphics[width=\columnwidth]{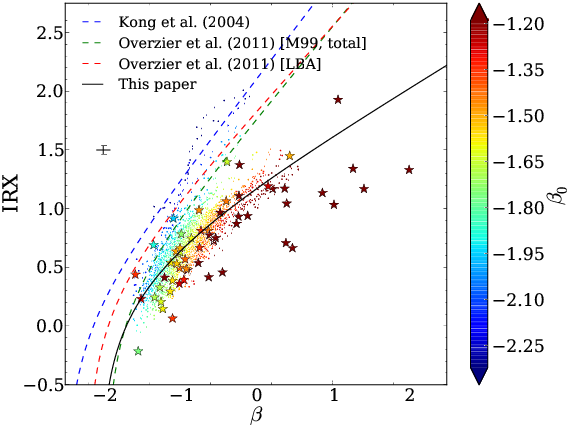}
\caption{Same as Fig.~\ref{fig:irx-beta}. The star symbols represent the subsample of 58 galaxies selected from the HRS that have $\chi^2<5$. The colours of individual data points and entire galaxies correspond to the same value of $\beta_0$, allowing for a direct comparison.}
\label{fig:irx-beta-age}
\end{figure}

In Fig.~\ref{fig:irx-beta-age}, we have plotted the IRX--$\beta$ diagram, the colour of each galaxy representing the value of $\beta_0$.
We see that the range of IRX covered by entire galaxies is similar to the one for individual regions. However, some HRS galaxies have a particularly high $\beta$, with 19\% of them having $\beta>0$, compared to 2\% for individual regions. The trend between $\beta_0$ and the perpendicular distance that was found inside galaxies can also be retrieved for entire galaxies. Galaxies that have a low $\beta_0$ are close to the starburst IRX--$\beta$ relation, and $\beta_0$ increases as a function of the perpendicular distance from the starburst IRX--$\beta$ relation. We see that a number of galaxies also have a particularly red UV slope with $\beta_0>-1.2$. Close inspection shows that many of these galaxies tend to be HI--deficient, with the HI deficiency computed as the logarithmic difference between the observed and the expected HI mass \citep{haynes1984a}. Their extremely red colour is most likely due to star formation that has been quenched because of a lack of a gas reservoir to feed from \citep{boselli2006b}. We find that there is a correlation between $\beta_0$ and the HI deficiency, with $\rho=0.68$, more gas--rich galaxies having a bluer $\beta_0$.

Galaxies that have a normal HI content (HI deficiency lower than 0.4) have $<\beta_0>=-1.41\pm0.38$, whereas $<\beta_0>=-0.78\pm0.72$ for HI deficient ones. Such a dispersion necessarily involves important variations in the recent SFH. HI deficient galaxies may have lost a large fraction of their gas over a relatively short time scale due to ram pressure for instance \citep{boselli2006a,boselli2008a}. Cutting abruptly star formation in such a way can easily create extreme values of $\beta_0$. This is in agreement with the result from \cite{cortese2008a} who showed that standard attenuation correction recipes fail for HI deficient galaxies. In the case of non--deficient galaxies, in addition to systematic errors, there are hints that it could be due to important variations of the SFH as shown by \cite{boselli2001a,gavazzi2002a,boselli2009a}, with nearly an order of magnitude of variation of the birthrate parameter at a given mass for a sample of normal star--forming galaxies.

\subsection{The variation of the UV slope and the standard stellar age estimators}

\begin{figure*}[!htbp]
\centering
\includegraphics[width=\columnwidth]{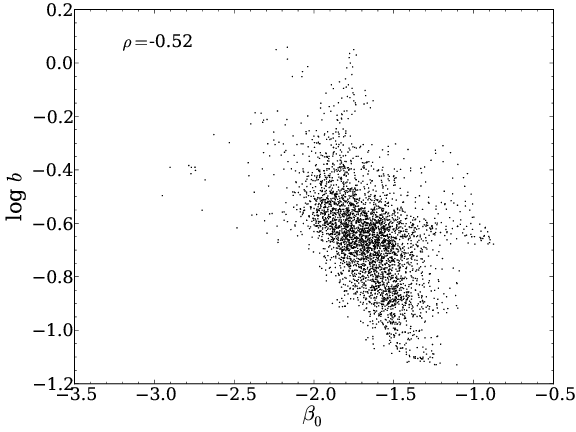}
\includegraphics[width=\columnwidth]{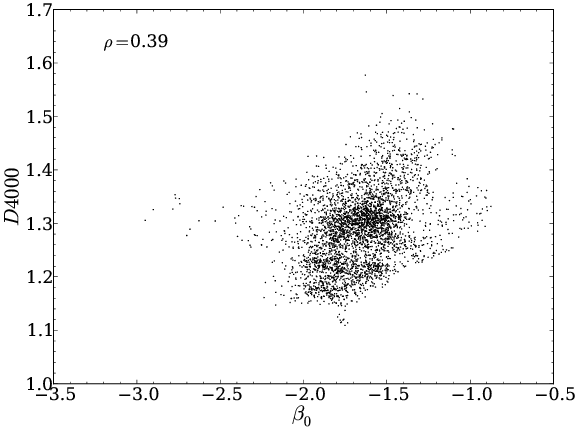}
\caption{Relation between $\beta_0$ and the birthrate parameter (left), and the $D4000$ parameter (right). The Spearman correlation coefficient $\rho$ is indicated on the top left hand corner of each figure.}
\label{fig:correl-beta0-age}
\end{figure*}

The age of the stellar populations has long been suspected to be the reason why star--forming galaxies deviate from the standard starburst relations. As mentioned earlier, the birthrate parameter $b$, the H$\alpha$ equivalent width, the $U-B$ colour, or the $D4000$ index have routinely been used to estimate the age. However neither of these parameters give a good, direct estimate of $\beta_0$, but they are affected by different specific biases. In Fig.~\ref{fig:correl-beta0-age} we present the relations between $\beta_0$, $D4000$, and the birthrate parameter.

It turns out that neither $b$ nor the $D4000$ index are good predictors of $\beta_0$. The Spearman correlation coefficient is $\rho=-0.52$ between $\log b$ and $\beta_0$, and $\rho=0.39$ between the $D4000$ index and $\beta_0$. As shown in \cite{boquien2009b}, this is likely due to mixing of successive generations of stellar populations. Indeed, these indicators are more accurate in case of a single, instantaneous starburst which prevents contamination by other populations. Therefore the usual age estimators tend to provide biased results as not all the emission is due to a recent instantaneous burst of star formation. The birthrate parameter is sensitive to recent star formation as well as the average star formation over the lifetime of the galaxy, which is closely linked to near--IR emission but mostly unrelated to UV. Conversely, the $D4000$ index (or the $U-B$ colour which straddles across the Balmer break), is sensitive to star formation over a longer timescale than star formation tracers, taking into account populations that do not emit significantly in the UV anymore. Simple age estimators are therefore not good estimators of $\beta_0$.

\section{Conclusion\label{sec:conclusion}}

In order to understand why star--forming spirals deviate from the starburst IRX--$\beta$ relation, we have modelled from the FUV to the FIR, including optical and near--IR data, a sample of 7 nearby, reasonably face--on, star--forming spirals drawn from the HRS. We have used the CIGALE code to estimate a large number of physical parameters on a pixel--by--pixel basis in each galaxy. The main results are:

\begin{itemize}
 \item The use of a canonical starburst $A_{FUV}$--$\beta$ relation on normal star--forming galaxies may produce an overestimate of the SFR by almost an order of magnitude, severely hampering the evaluation of the contribution of these galaxies to the cosmic star formation.
 \item The deviation from the starburst relation cannot be explained in terms of the stellar mass, the bolometric or dust luminosity, the SFR (either instantaneous or averaged over 10~Myr or 100~Myr), $D4000$, the birthrate parameter, or the mass--normalised age.
 \item The deviation from the starburst relation is found to be primarily due to significant variations in the intrinsic UV slope $\beta_0$ from one region to another. A variation of the slope $\delta$ of the attenuation curve may also play a secondary role.
 \item Even though the slope of the attenuation law is left unconstrained by CIGALE, data points with a low value of $\beta_0$ tend to have a starburst--like attenuation curve whereas regions that have a larger $\beta_0$ tend to have a steeper attenuation curve.
 \item New $A_{FUV}$--$\beta$ and $A_{FUV}$--IRX relations are provided. While they provide statistically accurate estimates of the attenuation, due to variations of $\beta_0$ and $\delta$ from one region to another, they are not physically motivated.
 \item The results obtained on a sample of star--forming spirals from the HRS are consistent with the ones obtained on subregions.
\end{itemize}

These results confirm that $\beta$ is not a good tracer of dust attenuation in normal star--forming galaxies if variations of the attenuation law and especially $\beta_0$ are not taken into account. As such great care must be used when such relations are applied on normal star--forming galaxies without proper modelling.

\begin{acknowledgements}
We thank the referee for useful comments that have helped improve the manuscript.

MB thanks S. Boissier and \'E. Giovannoli for enlightening discussions.

{\it Herschel} is an ESA space observatory with science instruments provided by European-led Principal Investigator consortia and with important participation from NASA.

SPIRE has been developed by a consortium of institutes led by Cardiff University (UK) and including Univ. Lethbridge (Canada); NAOC (China); CEA, LAM (France); IFSI, Univ. Padua (Italy); IAC (Spain); Stockholm Observatory (Sweden); Imperial College London, RAL, UCL-MSSL, UKATC, Univ. Sussex (UK); and Caltech, JPL, NHSC, Univ. Colorado (USA). This development has been supported by national funding agencies: CSA (Canada); NAOC (China); CEA, CNES, CNRS (France); ASI (Italy); MCINN (Spain); SNSB (Sweden); STFC (UK); and NASA (USA).

This research has made use of the NASA/IPAC Infrared Science Archive, which is operated by the Jet Propulsion Laboratory, California Institute of Technology, under contract with the National Aeronautics and Space Administration.

The research leading to these results has received funding from the European Community's Seventh Framework Programme (/FP7/2007-2013/) under grant agreement No 229517.
\end{acknowledgements}

\bibliographystyle{aa}
\bibliography{article}

\begin{thebibliography}{84}
\expandafter\ifx\csname natexlab\endcsname\relax\def\natexlab#1{#1}\fi

\bibitem[{{Abazajian} {et~al.}(2009){Abazajian}, {Adelman-McCarthy},
  {Ag{\"u}eros}, {Allam}, {Allende Prieto}, {An}, {Anderson}, {Anderson},
  {Annis}, {Bahcall}, \& et~al.}]{abazajian2009a}
{Abazajian}, K.~N., {Adelman-McCarthy}, J.~K., {Ag{\"u}eros}, M.~A., {et~al.}
  2009, \apjs, 182, 543

\bibitem[{{Anders} \& {Fritze-v.~Alvensleben}(2003)}]{anders2003a}
{Anders}, P. \& {Fritze-v.~Alvensleben}, U. 2003, \aap, 401, 1063

\bibitem[{{Aniano} {et~al.}(2011){Aniano}, {Draine}, {Gordon}, \&
  {Sandstrom}}]{aniano2011a}
{Aniano}, G., {Draine}, B.~T., {Gordon}, K.~D., \& {Sandstrom}, K. 2011, \pasp,
  123, 1218

\bibitem[{{Barnes} \& {Hernquist}(1992)}]{barnes1992a}
{Barnes}, J.~E. \& {Hernquist}, L. 1992, \nat, 360, 715

\bibitem[{{Bell}(2002)}]{bell2002a}
{Bell}, E.~F. 2002, \apj, 577, 150

\bibitem[{{Bendo} {et~al.}(2012, submitted){Bendo}, {Galliano}, \&
  {Madden}}]{bendo2012a}
{Bendo}, G., {Galliano}, F., \& {Madden}, S. 2012, submitted, \mnras

\bibitem[{{Boissier} {et~al.}(2007){Boissier}, {Gil de Paz}, {Boselli},
  {Madore}, {Buat}, {Cortese}, {Burgarella}, {Mu{\~n}oz-Mateos}, {Barlow},
  {Forster}, {Friedman}, {Martin}, {Morrissey}, {Neff}, {Schiminovich},
  {Seibert}, {Small}, {Wyder}, {Bianchi}, {Donas}, {Heckman}, {Lee},
  {Milliard}, {Rich}, {Szalay}, {Welsh}, \& {Yi}}]{boissier2007a}
{Boissier}, S., {Gil de Paz}, A., {Boselli}, A., {et~al.} 2007, \apjs, 173, 524

\bibitem[{{Boquien} {et~al.}(2011){Boquien}, {Calzetti}, {Combes}, {Henkel},
  {Israel}, {Kramer}, {Rela{\~n}o}, {Verley}, {van der Werf}, {Xilouris}, \&
  {The HERM33ES Team}}]{boquien2011a}
{Boquien}, M., {Calzetti}, D., {Combes}, F., {et~al.} 2011, \aj, 142, 111

\bibitem[{{Boquien} {et~al.}(2009){Boquien}, {Calzetti}, {Kennicutt}, {Dale},
  {Engelbracht}, {Gordon}, {Hong}, {Lee}, \& {Portouw}}]{boquien2009b}
{Boquien}, M., {Calzetti}, D., {Kennicutt}, R., {et~al.} 2009, \apj, 706, 553

\bibitem[{{Boquien} {et~al.}(2010){Boquien}, {Duc}, {Galliano}, {Braine},
  {Lisenfeld}, {Charmandaris}, \& {Appleton}}]{boquien2010c}
{Boquien}, M., {Duc}, P., {Galliano}, F., {et~al.} 2010, \aj, 140, 2124

\bibitem[{{Boselli} {et~al.}(2009){Boselli}, {Boissier}, {Cortese}, {Buat},
  {Hughes}, \& {Gavazzi}}]{boselli2009a}
{Boselli}, A., {Boissier}, S., {Cortese}, L., {et~al.} 2009, \apj, 706, 1527

\bibitem[{{Boselli} {et~al.}(2008){Boselli}, {Boissier}, {Cortese}, \&
  {Gavazzi}}]{boselli2008a}
{Boselli}, A., {Boissier}, S., {Cortese}, L., \& {Gavazzi}, G. 2008, \apj, 674,
  742

\bibitem[{{Boselli} {et~al.}(2006){Boselli}, {Boissier}, {Cortese}, {Gil de
  Paz}, {Seibert}, {Madore}, {Buat}, \& {Martin}}]{boselli2006a}
{Boselli}, A., {Boissier}, S., {Cortese}, L., {et~al.} 2006, \apj, 651, 811

\bibitem[{{Boselli} {et~al.}(2010){Boselli}, {Eales}, {Cortese}, {Bendo},
  {Chanial}, {Buat}, {Davies}, {Auld}, {Rigby}, {Baes}, {Barlow}, {Bock},
  {Bradford}, {Castro-Rodriguez}, {Charlot}, {Clements}, {Cormier}, {Dwek},
  {Elbaz}, {Galametz}, {Galliano}, {Gear}, {Glenn}, {Gomez}, {Griffin}, {Hony},
  {Isaak}, {Levenson}, {Lu}, {Madden}, {O'Halloran}, {Okamura}, {Oliver},
  {Page}, {Panuzzo}, {Papageorgiou}, {Parkin}, {Perez-Fournon}, {Pohlen},
  {Rangwala}, {Roussel}, {Rykala}, {Sacchi}, {Sauvage}, {Schulz}, {Schirm},
  {Smith}, {Spinoglio}, {Stevens}, {Symeonidis}, {Vaccari}, {Vigroux},
  {Wilson}, {Wozniak}, {Wright}, \& {Zeilinger}}]{boselli2010a}
{Boselli}, A., {Eales}, S., {Cortese}, L., {et~al.} 2010, \pasp, 122, 261

\bibitem[{{Boselli} \& {Gavazzi}(2006)}]{boselli2006b}
{Boselli}, A. \& {Gavazzi}, G. 2006, \pasp, 118, 517

\bibitem[{{Boselli} {et~al.}(2001){Boselli}, {Gavazzi}, {Donas}, \&
  {Scodeggio}}]{boselli2001a}
{Boselli}, A., {Gavazzi}, G., {Donas}, J., \& {Scodeggio}, M. 2001, \aj, 121,
  753

\bibitem[{{Bouwens} {et~al.}(2009){Bouwens}, {Illingworth}, {Franx}, {Chary},
  {Meurer}, {Conselice}, {Ford}, {Giavalisco}, \& {van Dokkum}}]{bouwens2009a}
{Bouwens}, R.~J., {Illingworth}, G.~D., {Franx}, M., {et~al.} 2009, \apj, 705,
  936

\bibitem[{{Buat} {et~al.}(2002){Buat}, {Boselli}, {Gavazzi}, \&
  {Bonfanti}}]{buat2002a}
{Buat}, V., {Boselli}, A., {Gavazzi}, G., \& {Bonfanti}, C. 2002, \aap, 383,
  801

\bibitem[{{Buat} {et~al.}(2011{\natexlab{a}}){Buat}, {Giovannoli}, {Heinis},
  {Charmandaris}, {Coia}, {Daddi}, {Dickinson}, {Elbaz}, {Hwang}, {Morrison},
  {Dasyra}, {Aussel}, {Altieri}, {Dannerbauer}, {Kartaltepe}, {Leiton},
  {Magdis}, {Magnelli}, \& {Popesso}}]{buat2011b}
{Buat}, V., {Giovannoli}, E., {Heinis}, S., {et~al.} 2011{\natexlab{a}}, \aap,
  533, A93+

\bibitem[{{Buat} {et~al.}(2011{\natexlab{b}}){Buat}, {Giovannoli}, {Takeuchi},
  {Heinis}, {Yuan}, {Burgarella}, {Noll}, \& {Iglesias-P{\'a}ramo}}]{buat2011a}
{Buat}, V., {Giovannoli}, E., {Takeuchi}, T.~T., {et~al.} 2011{\natexlab{b}},
  \aap, 529, A22+

\bibitem[{{Buat} {et~al.}(2005){Buat}, {Iglesias-P{\'a}ramo}, {Seibert},
  {Burgarella}, {Charlot}, {Martin}, {Xu}, {Heckman}, {Boissier}, {Boselli},
  {Barlow}, {Bianchi}, {Byun}, {Donas}, {Forster}, {Friedman}, {Jelinski},
  {Lee}, {Madore}, {Malina}, {Milliard}, {Morissey}, {Neff}, {Rich},
  {Schiminovitch}, {Siegmund}, {Small}, {Szalay}, {Welsh}, \&
  {Wyder}}]{buat2005a}
{Buat}, V., {Iglesias-P{\'a}ramo}, J., {Seibert}, M., {et~al.} 2005, \apjl,
  619, L51

\bibitem[{{Burgarella} {et~al.}(2005){Burgarella}, {Buat}, \&
  {Iglesias-P{\'a}ramo}}]{burgarella2005a}
{Burgarella}, D., {Buat}, V., \& {Iglesias-P{\'a}ramo}, J. 2005, \mnras, 360,
  1413

\bibitem[{{Burgarella} {et~al.}(2011){Burgarella}, {Heinis}, {Magdis}, {Auld},
  {Blain}, {Bock}, {Brisbin}, {Buat}, {Chanial}, {Clements}, {Cooray}, {Eales},
  {Franceschini}, {Giovannoli}, {Glenn}, {Gonz{\'a}lez Solares}, {Griffin},
  {Hwang}, {Ilbert}, {Marchetti}, {Mortier}, {Oliver}, {Page}, {Papageorgiou},
  {Pearson}, {P{\'e}rez-Fournon}, {Pohlen}, {Rawlings}, {Raymond},
  {Rigopoulou}, {Rodighiero}, {Roseboom}, {Rowan-Robinson}, {Scott}, {Seymour},
  {Smith}, {Symeonidis}, {Tugwell}, {Vaccari}, {Vieira}, {Viero}, {Vigroux},
  {Wang}, \& {Wright}}]{burgarella2011a}
{Burgarella}, D., {Heinis}, S., {Magdis}, G., {et~al.} 2011, \apjl, 734, L12

\bibitem[{{Calzetti} {et~al.}(2000){Calzetti}, {Armus}, {Bohlin}, {Kinney},
  {Koornneef}, \& {Storchi-Bergmann}}]{calzetti2000a}
{Calzetti}, D., {Armus}, L., {Bohlin}, R.~C., {et~al.} 2000, \apj, 533, 682

\bibitem[{{Calzetti} {et~al.}(2005){Calzetti}, {Kennicutt}, {Bianchi},
  {Thilker}, {Dale}, {Engelbracht}, {Leitherer}, {Meyer}, {Sosey}, {Mutchler},
  {Regan}, {Thornley}, {Armus}, {Bendo}, {Boissier}, {Boselli}, {Draine},
  {Gordon}, {Helou}, {Hollenbach}, {Kewley}, {Madore}, {Martin}, {Murphy},
  {Rieke}, {Rieke}, {Roussel}, {Sheth}, {Smith}, {Walter}, {White}, {Yi},
  {Scoville}, {Polletta}, \& {Lindler}}]{calzetti2005a}
{Calzetti}, D., {Kennicutt}, R.~C., {Bianchi}, L., {et~al.} 2005, \apj, 633,
  871

\bibitem[{{Calzetti} {et~al.}(2007){Calzetti}, {Kennicutt}, {Engelbracht},
  {Leitherer}, {Draine}, {Kewley}, {Moustakas}, {Sosey}, {Dale}, {Gordon},
  {Helou}, {Hollenbach}, {Armus}, {Bendo}, {Bot}, {Buckalew}, {Jarrett}, {Li},
  {Meyer}, {Murphy}, {Prescott}, {Regan}, {Rieke}, {Roussel}, {Sheth}, {Smith},
  {Thornley}, \& {Walter}}]{calzetti2007a}
{Calzetti}, D., {Kennicutt}, R.~C., {Engelbracht}, C.~W., {et~al.} 2007, \apj,
  666, 870

\bibitem[{{Calzetti} {et~al.}(1994){Calzetti}, {Kinney}, \&
  {Storchi-Bergmann}}]{calzetti1994a}
{Calzetti}, D., {Kinney}, A.~L., \& {Storchi-Bergmann}, T. 1994, \apj, 429, 582

\bibitem[{{Cardelli} {et~al.}(1989){Cardelli}, {Clayton}, \&
  {Mathis}}]{cardelli1989a}
{Cardelli}, J.~A., {Clayton}, G.~C., \& {Mathis}, J.~S. 1989, \apj, 345, 245

\bibitem[{{Charlot} \& {Fall}(2000)}]{charlot2000a}
{Charlot}, S. \& {Fall}, S.~M. 2000, \apj, 539, 718

\bibitem[{{Chary} \& {Elbaz}(2001)}]{chary2001a}
{Chary}, R. \& {Elbaz}, D. 2001, \apj, 556, 562

\bibitem[{{Ciesla} {et~al.}(2012, in preparation){Ciesla}, {Boselli}, \&
  {Bendo}}]{ciesla2012a}
{Ciesla}, L., {Boselli}, A., \& {Bendo}, G. 2012, in preparation, \aap

\bibitem[{{Conroy} {et~al.}(2010){Conroy}, {Schiminovich}, \&
  {Blanton}}]{conroy2010a}
{Conroy}, C., {Schiminovich}, D., \& {Blanton}, M.~R. 2010, \apj, 718, 184

\bibitem[{{Cortese} {et~al.}(2006){Cortese}, {Boselli}, {Buat}, {Gavazzi},
  {Boissier}, {Gil de Paz}, {Seibert}, {Madore}, \& {Martin}}]{cortese2006b}
{Cortese}, L., {Boselli}, A., {Buat}, V., {et~al.} 2006, \apj, 637, 242

\bibitem[{{Cortese} {et~al.}(2008){Cortese}, {Boselli}, {Franzetti}, {Decarli},
  {Gavazzi}, {Boissier}, \& {Buat}}]{cortese2008a}
{Cortese}, L., {Boselli}, A., {Franzetti}, P., {et~al.} 2008, \mnras, 386, 1157

\bibitem[{{Dale} {et~al.}(2007){Dale}, {Gil de Paz}, {Gordon}, {Hanson},
  {Armus}, {Bendo}, {Bianchi}, {Block}, {Boissier}, {Boselli}, {Buckalew},
  {Buat}, {Burgarella}, {Calzetti}, {Cannon}, {Engelbracht}, {Helou},
  {Hollenbach}, {Jarrett}, {Kennicutt}, {Leitherer}, {Li}, {Madore}, {Martin},
  {Meyer}, {Murphy}, {Regan}, {Roussel}, {Smith}, {Sosey}, {Thilker}, \&
  {Walter}}]{dale2007a}
{Dale}, D.~A., {Gil de Paz}, A., {Gordon}, K.~D., {et~al.} 2007, \apj, 655, 863

\bibitem[{{Dale} \& {Helou}(2002)}]{dale2002a}
{Dale}, D.~A. \& {Helou}, G. 2002, \apj, 576, 159

\bibitem[{{Dav{\'e}} {et~al.}(2011){Dav{\'e}}, {Oppenheimer}, \&
  {Finlator}}]{dave2011a}
{Dav{\'e}}, R., {Oppenheimer}, B.~D., \& {Finlator}, K. 2011, \mnras, 415, 11

\bibitem[{{Gavazzi} {et~al.}(2002){Gavazzi}, {Boselli}, {Pedotti}, {Gallazzi},
  \& {Carrasco}}]{gavazzi2002a}
{Gavazzi}, G., {Boselli}, A., {Pedotti}, P., {Gallazzi}, A., \& {Carrasco}, L.
  2002, \aap, 396, 449

\bibitem[{{Gil de Paz} {et~al.}(2007){Gil de Paz}, {Boissier}, {Madore},
  {Seibert}, {Joe}, {Boselli}, {Wyder}, {Thilker}, {Bianchi}, {Rey}, {Rich},
  {Barlow}, {Conrow}, {Forster}, {Friedman}, {Martin}, {Morrissey}, {Neff},
  {Schiminovich}, {Small}, {Donas}, {Heckman}, {Lee}, {Milliard}, {Szalay}, \&
  {Yi}}]{gildepaz2007a}
{Gil de Paz}, A., {Boissier}, S., {Madore}, B.~F., {et~al.} 2007, \apjs, 173,
  185

\bibitem[{{Giovannoli} {et~al.}(2011){Giovannoli}, {Buat}, {Noll},
  {Burgarella}, \& {Magnelli}}]{giovannoli2011a}
{Giovannoli}, E., {Buat}, V., {Noll}, S., {Burgarella}, D., \& {Magnelli}, B.
  2011, \aap, 525, A150+

\bibitem[{{Gordon} {et~al.}(2003){Gordon}, {Clayton}, {Misselt}, {Landolt}, \&
  {Wolff}}]{gordon2003a}
{Gordon}, K.~D., {Clayton}, G.~C., {Misselt}, K.~A., {Landolt}, A.~U., \&
  {Wolff}, M.~J. 2003, \apj, 594, 279

\bibitem[{{Gordon} {et~al.}(2004){Gordon}, {P{\'e}rez-Gonz{\'a}lez}, {Misselt},
  {Murphy}, {Bendo}, {Walter}, {Thornley}, {Kennicutt}, {Rieke}, {Engelbracht},
  {Smith}, {Alonso-Herrero}, {Appleton}, {Calzetti}, {Dale}, {Draine},
  {Frayer}, {Helou}, {Hinz}, {Hines}, {Kelly}, {Morrison}, {Muzerolle},
  {Regan}, {Stansberry}, {Stolovy}, {Storrie-Lombardi}, {Su}, \&
  {Young}}]{gordon2004a}
{Gordon}, K.~D., {P{\'e}rez-Gonz{\'a}lez}, P.~G., {Misselt}, K.~A., {et~al.}
  2004, \apjs, 154, 215

\bibitem[{{Griffin} {et~al.}(2010){Griffin}, {Abergel}, {Abreu}, {Ade},
  {Andr{\'e}}, {Augueres}, {Babbedge}, {Bae}, {Baillie}, {Baluteau}, {Barlow},
  {Bendo}, {Benielli}, {Bock}, {Bonhomme}, {Brisbin}, {Brockley-Blatt},
  {Caldwell}, {Cara}, {Castro-Rodriguez}, {Cerulli}, {Chanial}, {Chen},
  {Clark}, {Clements}, {Clerc}, {Coker}, {Communal}, {Conversi}, {Cox},
  {Crumb}, {Cunningham}, {Daly}, {Davis}, {de Antoni}, {Delderfield}, {Devin},
  {di Giorgio}, {Didschuns}, {Dohlen}, {Donati}, {Dowell}, {Dowell}, {Duband},
  {Dumaye}, {Emery}, {Ferlet}, {Ferrand}, {Fontignie}, {Fox}, {Franceschini},
  {Frerking}, {Fulton}, {Garcia}, {Gastaud}, {Gear}, {Glenn}, {Goizel},
  {Griffin}, {Grundy}, {Guest}, {Guillemet}, {Hargrave}, {Harwit}, {Hastings},
  {Hatziminaoglou}, {Herman}, {Hinde}, {Hristov}, {Huang}, {Imhof}, {Isaak},
  {Israelsson}, {Ivison}, {Jennings}, {Kiernan}, {King}, {Lange}, {Latter},
  {Laurent}, {Laurent}, {Leeks}, {Lellouch}, {Levenson}, {Li}, {Li},
  {Lilienthal}, {Lim}, {Liu}, {Lu}, {Madden}, {Mainetti}, {Marliani}, {McKay},
  {Mercier}, {Molinari}, {Morris}, {Moseley}, {Mulder}, {Mur}, {Naylor},
  {Nguyen}, {O'Halloran}, {Oliver}, {Olofsson}, {Olofsson}, {Orfei}, {Page},
  {Pain}, {Panuzzo}, {Papageorgiou}, {Parks}, {Parr-Burman}, {Pearce},
  {Pearson}, {P{\'e}rez-Fournon}, {Pinsard}, {Pisano}, {Podosek}, {Pohlen},
  {Polehampton}, {Pouliquen}, {Rigopoulou}, {Rizzo}, {Roseboom}, {Roussel},
  {Rowan-Robinson}, {Rownd}, {Saraceno}, {Sauvage}, {Savage}, {Savini},
  {Sawyer}, {Scharmberg}, {Schmitt}, {Schneider}, {Schulz}, {Schwartz},
  {Shafer}, {Shupe}, {Sibthorpe}, {Sidher}, {Smith}, {Smith}, {Smith},
  {Spencer}, {Stobie}, {Sudiwala}, {Sukhatme}, {Surace}, {Stevens}, {Swinyard},
  {Trichas}, {Tourette}, {Triou}, {Tseng}, {Tucker}, {Turner}, {Vaccari},
  {Valtchanov}, {Vigroux}, {Virique}, {Voellmer}, {Walker}, {Ward}, {Waskett},
  {Weilert}, {Wesson}, {White}, {Whitehouse}, {Wilson}, {Winter}, {Woodcraft},
  {Wright}, {Xu}, {Zavagno}, {Zemcov}, {Zhang}, \& {Zonca}}]{griffin2010a}
{Griffin}, M.~J., {Abergel}, A., {Abreu}, A., {et~al.} 2010, \aap, 518, L3

\bibitem[{{Hao} {et~al.}(2011){Hao}, {Kennicutt}, {Johnson}, {Calzetti},
  {Dale}, \& {Moustakas}}]{hao2011a}
{Hao}, C.-N., {Kennicutt}, R.~C., {Johnson}, B.~D., {et~al.} 2011, \apj, 741,
  124

\bibitem[{{Haynes} \& {Giovanelli}(1984)}]{haynes1984a}
{Haynes}, M.~P. \& {Giovanelli}, R. 1984, \aj, 89, 758

\bibitem[{{Hopkins}(2004)}]{hopkins2004a}
{Hopkins}, A.~M. 2004, \apj, 615, 209

\bibitem[{{Johnson} {et~al.}(2007){Johnson}, {Schiminovich}, {Seibert},
  {Treyer}, {Martin}, {Barlow}, {Forster}, {Friedman}, {Morrissey}, {Neff},
  {Small}, {Wyder}, {Bianchi}, {Donas}, {Heckman}, {Lee}, {Madore}, {Milliard},
  {Rich}, {Szalay}, {Welsh}, \& {Yi}}]{johnson2007a}
{Johnson}, B.~D., {Schiminovich}, D., {Seibert}, M., {et~al.} 2007, \apjs, 173,
  392

\bibitem[{{Kauffmann} {et~al.}(2003){Kauffmann}, {Heckman}, {White}, {Charlot},
  {Tremonti}, {Brinchmann}, {Bruzual}, {Peng}, {Seibert}, {Bernardi},
  {Blanton}, {Brinkmann}, {Castander}, {Cs{\'a}bai}, {Fukugita}, {Ivezic},
  {Munn}, {Nichol}, {Padmanabhan}, {Thakar}, {Weinberg}, \&
  {York}}]{kauffmann2003a}
{Kauffmann}, G., {Heckman}, T.~M., {White}, S.~D.~M., {et~al.} 2003, \mnras,
  341, 33

\bibitem[{{Kennicutt} {et~al.}(2009){Kennicutt}, {Hao}, {Calzetti},
  {Moustakas}, {Dale}, {Bendo}, {Engelbracht}, {Johnson}, \&
  {Lee}}]{kennicutt2009a}
{Kennicutt}, R.~C., {Hao}, C., {Calzetti}, D., {et~al.} 2009, \apj, 703, 1672

\bibitem[{{Kewley} {et~al.}(2010){Kewley}, {Rupke}, {Jabran Zahid}, {Geller},
  \& {Barton}}]{kewley2010a}
{Kewley}, L.~J., {Rupke}, D., {Jabran Zahid}, H., {Geller}, M.~J., \& {Barton},
  E.~J. 2010, \apjl, 721, L48

\bibitem[{{Kitzbichler} \& {White}(2007)}]{kitzbichler2007a}
{Kitzbichler}, M.~G. \& {White}, S.~D.~M. 2007, \mnras, 376, 2

\bibitem[{{Kong} {et~al.}(2004){Kong}, {Charlot}, {Brinchmann}, \&
  {Fall}}]{kong2004a}
{Kong}, X., {Charlot}, S., {Brinchmann}, J., \& {Fall}, S.~M. 2004, \mnras,
  349, 769

\bibitem[{{Kramer} {et~al.}(2010){Kramer}, {Buchbender}, {Xilouris}, {Boquien},
  {Braine}, {Calzetti}, {Lord}, {Mookerjea}, {Quintana-Lacaci}, {Rela{\~n}o},
  {Stacey}, {Tabatabaei}, {Verley}, {Aalto}, {Akras}, {Albrecht}, {Anderl},
  {Beck}, {Bertoldi}, {Combes}, {Dumke}, {Garcia-Burillo}, {Gonzalez},
  {Gratier}, {G{\"u}sten}, {Henkel}, {Israel}, {Koribalski}, {Lundgren},
  {Martin-Pintado}, {R{\"o}llig}, {Rosolowsky}, {Schuster}, {Sheth}, {Sievers},
  {Stutzki}, {Tilanus}, {van der Tak}, {van der Werf}, \&
  {Wiedner}}]{kramer2010a}
{Kramer}, C., {Buchbender}, C., {Xilouris}, E.~M., {et~al.} 2010, \aap, 518,
  L67+

\bibitem[{{Leroy} {et~al.}(2008){Leroy}, {Walter}, {Brinks}, {Bigiel}, {de
  Blok}, {Madore}, \& {Thornley}}]{leroy2008a}
{Leroy}, A.~K., {Walter}, F., {Brinks}, E., {et~al.} 2008, \aj, 136, 2782

\bibitem[{{Madau} {et~al.}(1996){Madau}, {Ferguson}, {Dickinson}, {Giavalisco},
  {Steidel}, \& {Fruchter}}]{madau1996a}
{Madau}, P., {Ferguson}, H.~C., {Dickinson}, M.~E., {et~al.} 1996, \mnras, 283,
  1388

\bibitem[{{Magnelli} {et~al.}(2009){Magnelli}, {Elbaz}, {Chary}, {Dickinson},
  {Le Borgne}, {Frayer}, \& {Willmer}}]{magnelli2009a}
{Magnelli}, B., {Elbaz}, D., {Chary}, R.~R., {et~al.} 2009, \aap, 496, 57

\bibitem[{{Magnelli} {et~al.}(2011){Magnelli}, {Elbaz}, {Chary}, {Dickinson},
  {Le Borgne}, {Frayer}, \& {Willmer}}]{magnelli2011a}
{Magnelli}, B., {Elbaz}, D., {Chary}, R.~R., {et~al.} 2011, \aap, 528, A35+

\bibitem[{{Maraston}(2005)}]{maraston2005a}
{Maraston}, C. 2005, \mnras, 362, 799

\bibitem[{{Martin} {et~al.}(2005){Martin}, {Fanson}, {Schiminovich},
  {Morrissey}, {Friedman}, {Barlow}, {Conrow}, {Grange}, {Jelinsky},
  {Milliard}, {Siegmund}, {Bianchi}, {Byun}, {Donas}, {Forster}, {Heckman},
  {Lee}, {Madore}, {Malina}, {Neff}, {Rich}, {Small}, {Surber}, {Szalay},
  {Welsh}, \& {Wyder}}]{martin2005a}
{Martin}, D.~C., {Fanson}, J., {Schiminovich}, D., {et~al.} 2005, \apjl, 619,
  L1

\bibitem[{{Meurer} {et~al.}(1999){Meurer}, {Heckman}, \&
  {Calzetti}}]{meurer1999a}
{Meurer}, G.~R., {Heckman}, T.~M., \& {Calzetti}, D. 1999, \apj, 521, 64

\bibitem[{{Mu{\~n}oz-Mateos} {et~al.}(2009){Mu{\~n}oz-Mateos}, {Gil de Paz},
  {Boissier}, {Zamorano}, {Dale}, {P{\'e}rez-Gonz{\'a}lez}, {Gallego},
  {Madore}, {Bendo}, {Thornley}, {Draine}, {Boselli}, {Buat}, {Calzetti},
  {Moustakas}, \& {Kennicutt}}]{munoz2009a}
{Mu{\~n}oz-Mateos}, J.~C., {Gil de Paz}, A., {Boissier}, S., {et~al.} 2009,
  \apj, 701, 1965

\bibitem[{{Noll} {et~al.}(2009{\natexlab{a}}){Noll}, {Burgarella},
  {Giovannoli}, {Buat}, {Marcillac}, \& {Mu{\~n}oz-Mateos}}]{noll2009a}
{Noll}, S., {Burgarella}, D., {Giovannoli}, E., {et~al.} 2009{\natexlab{a}},
  \aap, 507, 1793

\bibitem[{{Noll} {et~al.}(2009{\natexlab{b}}){Noll}, {Pierini}, {Cimatti},
  {Daddi}, {Kurk}, {Bolzonella}, {Cassata}, {Halliday}, {Mignoli}, {Pozzetti},
  {Renzini}, {Berta}, {Dickinson}, {Franceschini}, {Rodighiero}, {Rosati}, \&
  {Zamorani}}]{noll2009b}
{Noll}, S., {Pierini}, D., {Cimatti}, A., {et~al.} 2009{\natexlab{b}}, \aap,
  499, 69

\bibitem[{{O'Donnell}(1994)}]{odonnell1994a}
{O'Donnell}, J.~E. 1994, \apj, 422, 158

\bibitem[{{Overzier} {et~al.}(2011){Overzier}, {Heckman}, {Wang}, {Armus},
  {Buat}, {Howell}, {Meurer}, {Seibert}, {Siana}, {Basu-Zych}, {Charlot},
  {Gon{\c c}alves}, {Martin}, {Neill}, {Rich}, {Salim}, \&
  {Schiminovich}}]{overzier2011a}
{Overzier}, R.~A., {Heckman}, T.~M., {Wang}, J., {et~al.} 2011, \apjl, 726, L7+

\bibitem[{{Panuzzo} {et~al.}(2007){Panuzzo}, {Granato}, {Buat}, {Inoue},
  {Silva}, {Iglesias-P{\'a}ramo}, \& {Bressan}}]{panuzzo2007a}
{Panuzzo}, P., {Granato}, G.~L., {Buat}, V., {et~al.} 2007, \mnras, 375, 640

\bibitem[{{P{\'e}rez-Gonz{\'a}lez} {et~al.}(2005){P{\'e}rez-Gonz{\'a}lez},
  {Rieke}, {Egami}, {Alonso-Herrero}, {Dole}, {Papovich}, {Blaylock}, {Jones},
  {Rieke}, {Rigby}, {Barmby}, {Fazio}, {Huang}, \& {Martin}}]{perez2005a}
{P{\'e}rez-Gonz{\'a}lez}, P.~G., {Rieke}, G.~H., {Egami}, E., {et~al.} 2005,
  \apj, 630, 82

\bibitem[{{Pilbratt} {et~al.}(2010){Pilbratt}, {Riedinger}, {Passvogel},
  {Crone}, {Doyle}, {Gageur}, {Heras}, {Jewell}, {Metcalfe}, {Ott}, \&
  {Schmidt}}]{pilbratt2010a}
{Pilbratt}, G.~L., {Riedinger}, J.~R., {Passvogel}, T., {et~al.} 2010, \aap,
  518, L1

\bibitem[{{Reddy} \& {Steidel}(2009)}]{reddy2009a}
{Reddy}, N.~A. \& {Steidel}, C.~C. 2009, \apj, 692, 778

\bibitem[{{Rieke} {et~al.}(2004){Rieke}, {Young}, {Engelbracht}, {Kelly},
  {Low}, {Haller}, {Beeman}, {Gordon}, {Stansberry}, {Misselt}, {Cadien},
  {Morrison}, {Rivlis}, {Latter}, {Noriega-Crespo}, {Padgett}, {Stapelfeldt},
  {Hines}, {Egami}, {Muzerolle}, {Alonso-Herrero}, {Blaylock}, {Dole}, {Hinz},
  {Le Floc'h}, {Papovich}, {P{\'e}rez-Gonz{\'a}lez}, {Smith}, {Su}, {Bennett},
  {Frayer}, {Henderson}, {Lu}, {Masci}, {Pesenson}, {Rebull}, {Rho}, {Keene},
  {Stolovy}, {Wachter}, {Wheaton}, {Werner}, \& {Richards}}]{rieke2004a}
{Rieke}, G.~H., {Young}, E.~T., {Engelbracht}, C.~W., {et~al.} 2004, \apjs,
  154, 25

\bibitem[{{Rodighiero} {et~al.}(2010){Rodighiero}, {Vaccari}, {Franceschini},
  {Tresse}, {Le Fevre}, {Le Brun}, {Mancini}, {Matute}, {Cimatti}, {Marchetti},
  {Ilbert}, {Arnouts}, {Bolzonella}, {Zucca}, {Bardelli}, {Lonsdale}, {Shupe},
  {Surace}, {Rowan-Robinson}, {Garilli}, {Zamorani}, {Pozzetti}, {Bondi}, {de
  la Torre}, {Vergani}, {Santini}, {Grazian}, \& {Fontana}}]{rodighiero2010a}
{Rodighiero}, G., {Vaccari}, M., {Franceschini}, A., {et~al.} 2010, \aap, 515,
  A8+

\bibitem[{{Rudolph} {et~al.}(2006){Rudolph}, {Fich}, {Bell}, {Norsen},
  {Simpson}, {Haas}, \& {Erickson}}]{rudolph2006a}
{Rudolph}, A.~L., {Fich}, M., {Bell}, G.~R., {et~al.} 2006, \apjs, 162, 346

\bibitem[{{Rupke} {et~al.}(2010{\natexlab{a}}){Rupke}, {Kewley}, \&
  {Barnes}}]{rupke2010a}
{Rupke}, D.~S.~N., {Kewley}, L.~J., \& {Barnes}, J.~E. 2010{\natexlab{a}},
  \apjl, 710, L156

\bibitem[{{Rupke} {et~al.}(2010{\natexlab{b}}){Rupke}, {Kewley}, \&
  {Chien}}]{rupke2010b}
{Rupke}, D.~S.~N., {Kewley}, L.~J., \& {Chien}, L.-H. 2010{\natexlab{b}}, \apj,
  723, 1255

\bibitem[{{Schiminovich} {et~al.}(2005){Schiminovich}, {Ilbert}, {Arnouts},
  {Milliard}, {Tresse}, {Le F{\`e}vre}, {Treyer}, {Wyder}, {Budav{\'a}ri},
  {Zucca}, {Zamorani}, {Martin}, {Adami}, {Arnaboldi}, {Bardelli}, {Barlow},
  {Bianchi}, {Bolzonella}, {Bottini}, {Byun}, {Cappi}, {Contini}, {Charlot},
  {Donas}, {Forster}, {Foucaud}, {Franzetti}, {Friedman}, {Garilli},
  {Gavignaud}, {Guzzo}, {Heckman}, {Hoopes}, {Iovino}, {Jelinsky}, {Le Brun},
  {Lee}, {Maccagni}, {Madore}, {Malina}, {Marano}, {Marinoni}, {McCracken},
  {Mazure}, {Meneux}, {Morrissey}, {Neff}, {Paltani}, {Pell{\`o}}, {Picat},
  {Pollo}, {Pozzetti}, {Radovich}, {Rich}, {Scaramella}, {Scodeggio},
  {Seibert}, {Siegmund}, {Small}, {Szalay}, {Vettolani}, {Welsh}, {Xu}, \&
  {Zanichelli}}]{schiminovich2005a}
{Schiminovich}, D., {Ilbert}, O., {Arnouts}, S., {et~al.} 2005, \apjl, 619, L47

\bibitem[{{Schlegel} {et~al.}(1998){Schlegel}, {Finkbeiner}, \&
  {Davis}}]{schlegel1998a}
{Schlegel}, D.~J., {Finkbeiner}, D.~P., \& {Davis}, M. 1998, \apj, 500, 525

\bibitem[{{Seibert} {et~al.}(2005){Seibert}, {Martin}, {Heckman}, {Buat},
  {Hoopes}, {Barlow}, {Bianchi}, {Byun}, {Donas}, {Forster}, {Friedman},
  {Jelinsky}, {Lee}, {Madore}, {Malina}, {Milliard}, {Morrissey}, {Neff},
  {Rich}, {Schiminovich}, {Siegmund}, {Small}, {Szalay}, {Welsh}, \&
  {Wyder}}]{seibert2005a}
{Seibert}, M., {Martin}, D.~C., {Heckman}, T.~M., {et~al.} 2005, \apjl, 619,
  L55

\bibitem[{{Skrutskie} {et~al.}(2006){Skrutskie}, {Cutri}, {Stiening},
  {Weinberg}, {Schneider}, {Carpenter}, {Beichman}, {Capps}, {Chester},
  {Elias}, {Huchra}, {Liebert}, {Lonsdale}, {Monet}, {Price}, {Seitzer},
  {Jarrett}, {Kirkpatrick}, {Gizis}, {Howard}, {Evans}, {Fowler}, {Fullmer},
  {Hurt}, {Light}, {Kopan}, {Marsh}, {McCallon}, {Tam}, {Van Dyk}, \&
  {Wheelock}}]{skrutskie2006a}
{Skrutskie}, M.~F., {Cutri}, R.~M., {Stiening}, R., {et~al.} 2006, \aj, 131,
  1163

\bibitem[{{Steidel} {et~al.}(1999){Steidel}, {Adelberger}, {Giavalisco},
  {Dickinson}, \& {Pettini}}]{steidel1999a}
{Steidel}, C.~C., {Adelberger}, K.~L., {Giavalisco}, M., {Dickinson}, M., \&
  {Pettini}, M. 1999, \apj, 519, 1

\bibitem[{{van der Burg} {et~al.}(2010){van der Burg}, {Hildebrandt}, \&
  {Erben}}]{vanderburg2010a}
{van der Burg}, R.~F.~J., {Hildebrandt}, H., \& {Erben}, T. 2010, \aap, 523,
  A74+

\bibitem[{{Walcher} {et~al.}(2008){Walcher}, {Lamareille}, {Vergani},
  {Arnouts}, {Buat}, {Charlot}, {Tresse}, {Le F{\`e}vre}, {Bolzonella},
  {Brinchmann}, {Pozzetti}, {Zamorani}, {Bottini}, {Garilli}, {Le Brun},
  {Maccagni}, {Milliard}, {Scaramella}, {Scodeggio}, {Vettolani}, {Zanichelli},
  {Adami}, {Bardelli}, {Cappi}, {Ciliegi}, {Contini}, {Franzetti}, {Foucaud},
  {Gavignaud}, {Guzzo}, {Ilbert}, {Iovino}, {McCracken}, {Marano}, {Marinoni},
  {Mazure}, {Meneux}, {Merighi}, {Paltani}, {Pell{\`o}}, {Pollo}, {Radovich},
  {Zucca}, {Lonsdale}, \& {Martin}}]{walcher2008a}
{Walcher}, C.~J., {Lamareille}, F., {Vergani}, D., {et~al.} 2008, \aap, 491,
  713

\bibitem[{{Wijesinghe} {et~al.}(2011){Wijesinghe}, {da Cunha}, {Hopkins},
  {Dunne}, {Sharp}, {Gunawardhana}, {Brough}, {Sadler}, {Driver}, {Baldry},
  {Bamford}, {Liske}, {Loveday}, {Norberg}, {Peacock}, {Popescu}, {Tuffs},
  {Andrae}, {Auld}, {Baes}, {Bland-Hawthorn}, {Buttiglione}, {Cava}, {Cameron},
  {Conselice}, {Cooray}, {Croom}, {Dariush}, {Dezotti}, {Dye}, {Eales},
  {Frenk}, {Fritz}, {Hill}, {Hopwood}, {Ibar}, {Ivison}, {Jarvis}, {Jones},
  {van Kampen}, {Kelvin}, {Kuijken}, {Maddox}, {Madore}, {Micha{\l}owski},
  {Nichol}, {Parkinson}, {Pascale}, {Pimbblet}, {Pohlen}, {Prescott},
  {Rhodighiero}, {Robotham}, {Rigby}, {Seibert}, {Sergeant}, {Smith}, {Temi},
  {Sutherland}, {Taylor}, {Thomas}, \& {van der Werf}}]{wijesinghe2011a}
{Wijesinghe}, D.~B., {da Cunha}, E., {Hopkins}, A.~M., {et~al.} 2011, \mnras,
  415, 1002

\bibitem[{{Wild} {et~al.}(2011){Wild}, {Charlot}, {Brinchmann}, {Heckman},
  {Vince}, {Pacifici}, \& {Chevallard}}]{wild2011a}
{Wild}, V., {Charlot}, S., {Brinchmann}, J., {et~al.} 2011, \mnras, 417, 1760

\bibitem[{{Witt} \& {Gordon}(2000)}]{witt2000a}
{Witt}, A.~N. \& {Gordon}, K.~D. 2000, \apj, 528, 799

\end{thebibliography}

\appendix
\section{Correlation between \texorpdfstring{$d_\perp$, $d_\parallel$}{the perpendicular distance, the parallel distance}, and the parameters}

\begin{figure*}[!htbp]
\centering
\includegraphics[width=0.6\columnwidth]{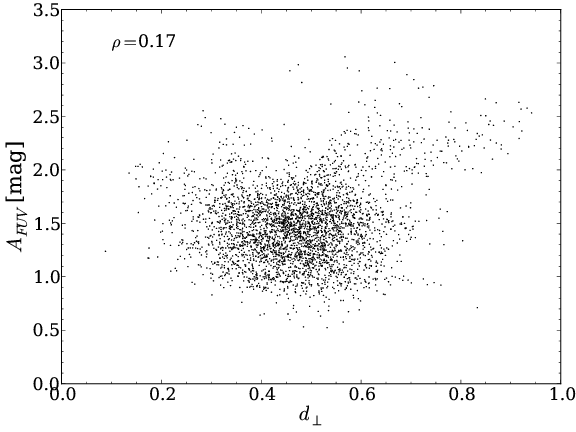}
\includegraphics[width=0.6\columnwidth]{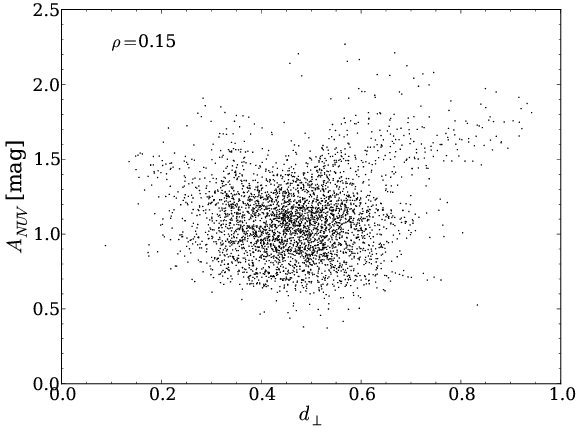}
\includegraphics[width=0.6\columnwidth]{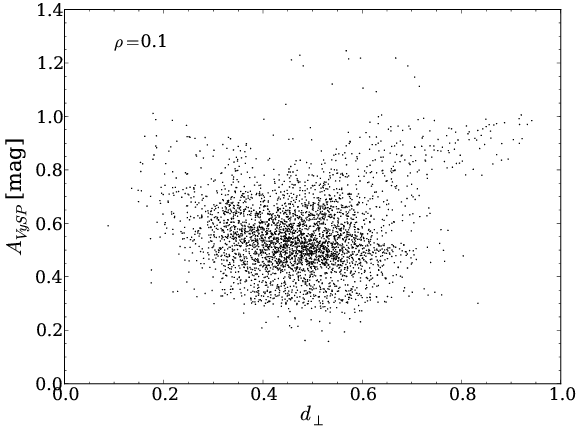}
\includegraphics[width=0.6\columnwidth]{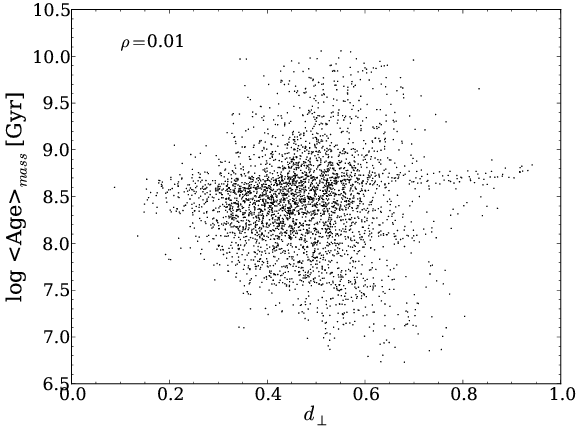}
\includegraphics[width=0.6\columnwidth]{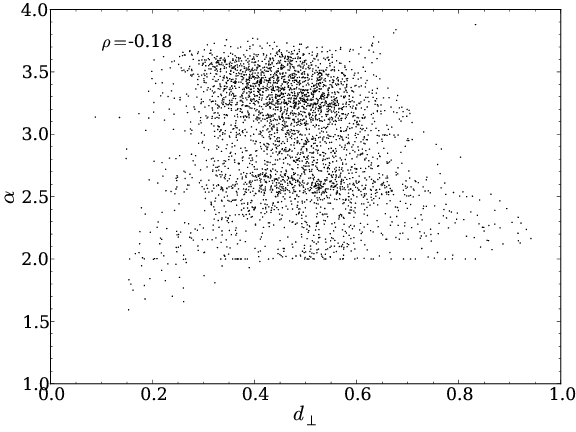}
\includegraphics[width=0.6\columnwidth]{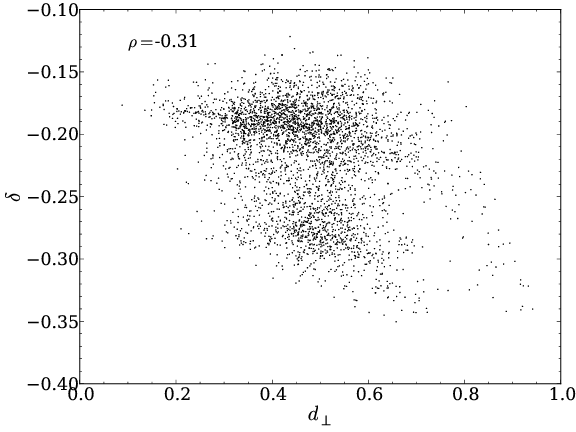}
\includegraphics[width=0.6\columnwidth]{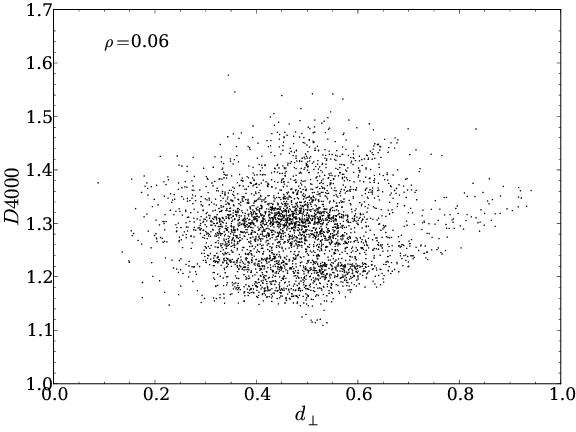}
\includegraphics[width=0.6\columnwidth]{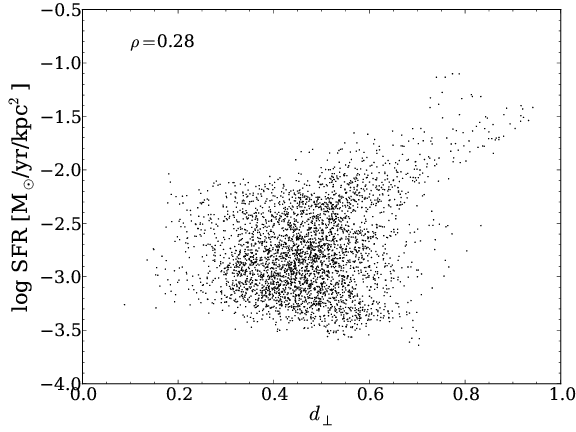}
\includegraphics[width=0.6\columnwidth]{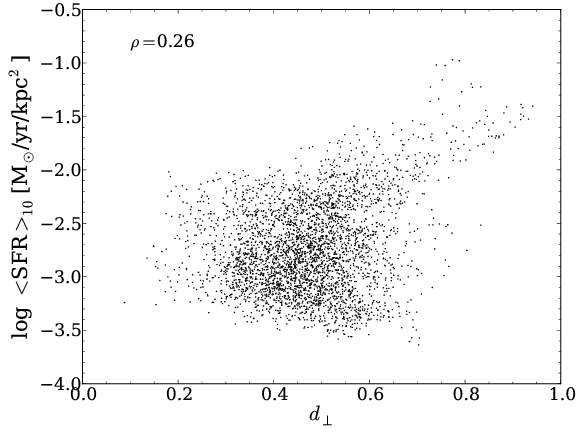}
\includegraphics[width=0.6\columnwidth]{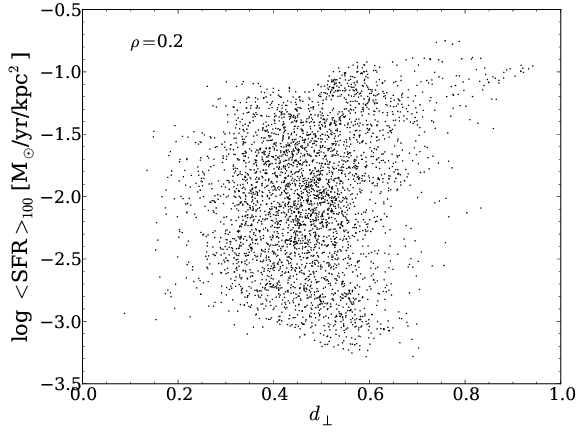}
\includegraphics[width=0.6\columnwidth]{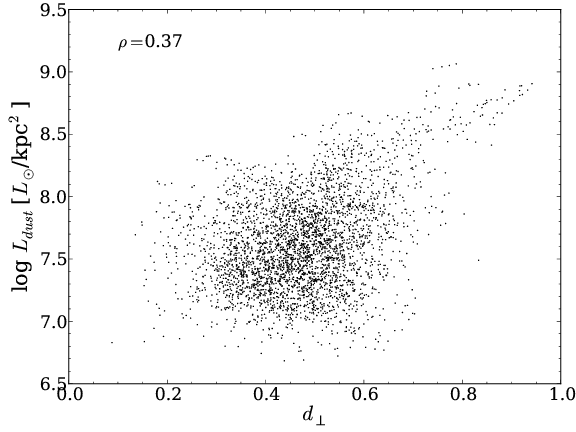}
\includegraphics[width=0.6\columnwidth]{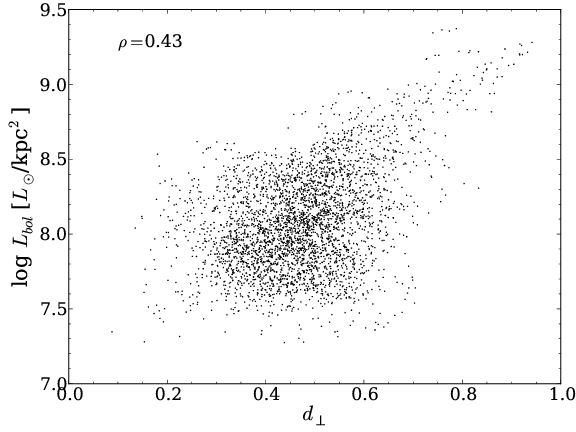}
\includegraphics[width=0.6\columnwidth]{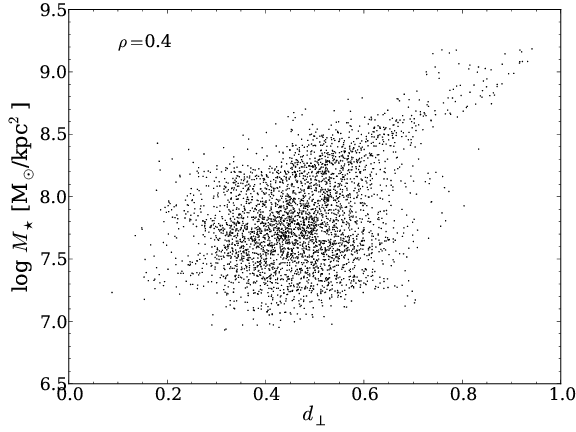}
\includegraphics[width=0.6\columnwidth]{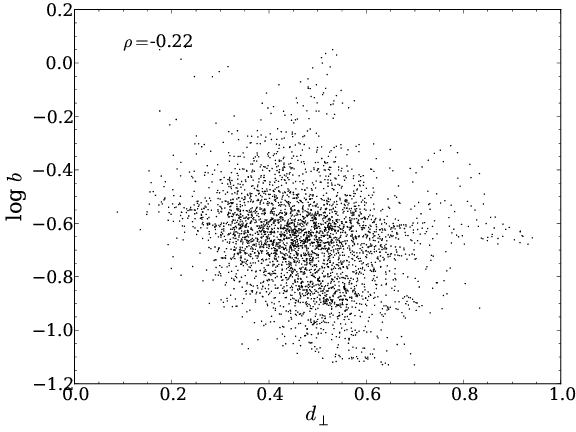}
\includegraphics[width=0.6\columnwidth]{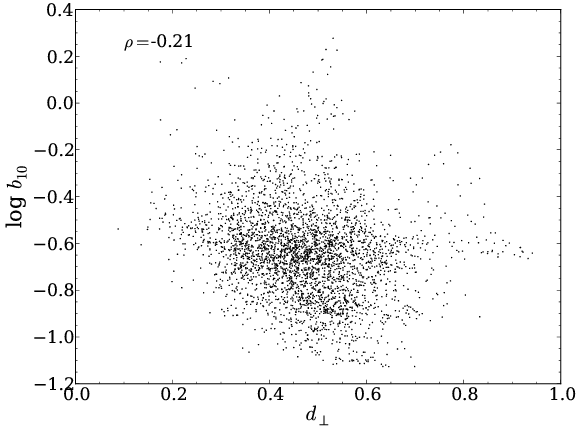}
\includegraphics[width=0.6\columnwidth]{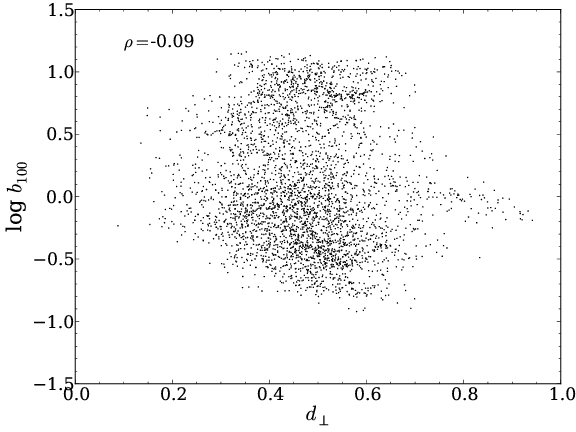}
\caption{Parameters versus $d_\perp$. The correlation coefficient is indicated on the top--left hand corner of each plot.}
\label{fig:dist-perp}
\end{figure*}

\begin{figure*}[!htbp]
\centering
\includegraphics[width=0.6\columnwidth]{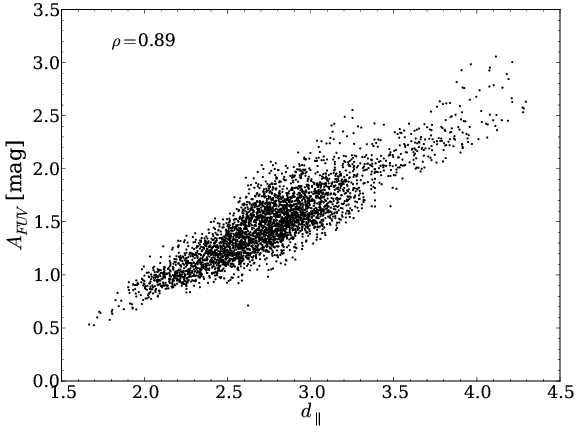}
\includegraphics[width=0.6\columnwidth]{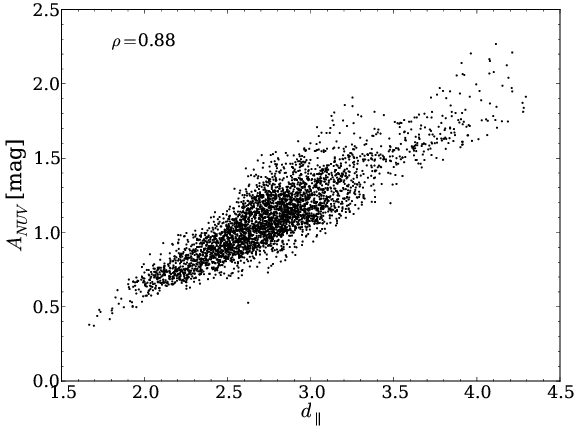}
\includegraphics[width=0.6\columnwidth]{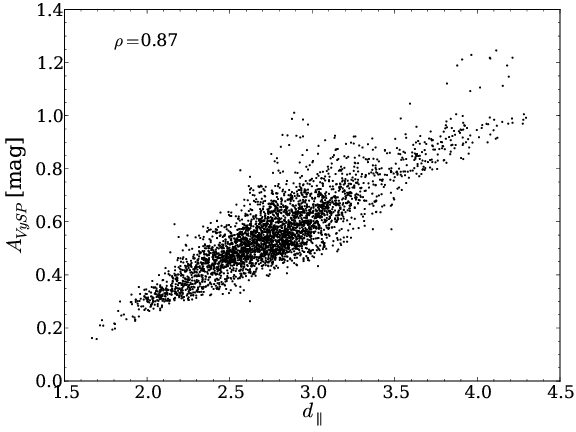}
\includegraphics[width=0.6\columnwidth]{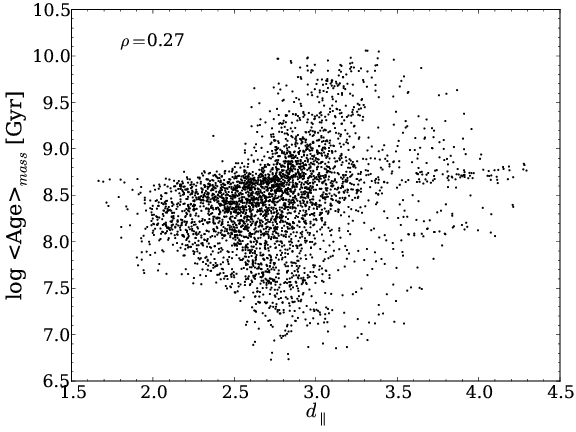}
\includegraphics[width=0.6\columnwidth]{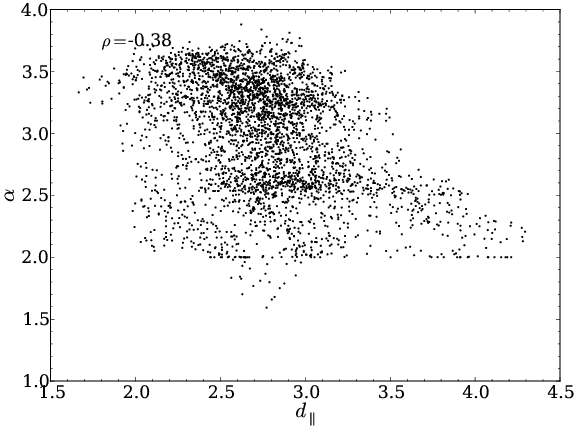}
\includegraphics[width=0.6\columnwidth]{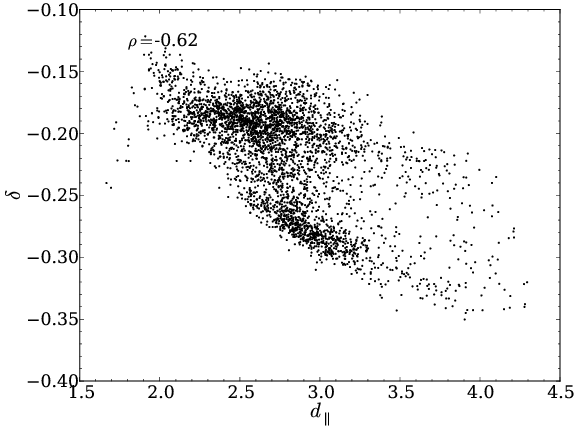}
\includegraphics[width=0.6\columnwidth]{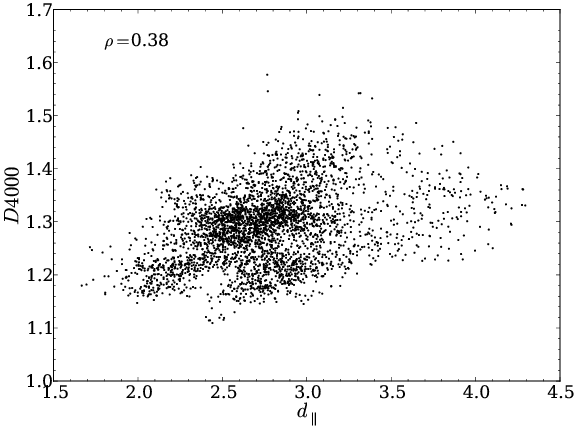}
\includegraphics[width=0.6\columnwidth]{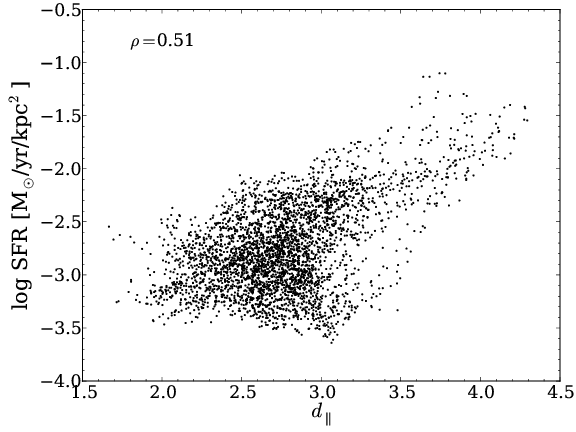}
\includegraphics[width=0.6\columnwidth]{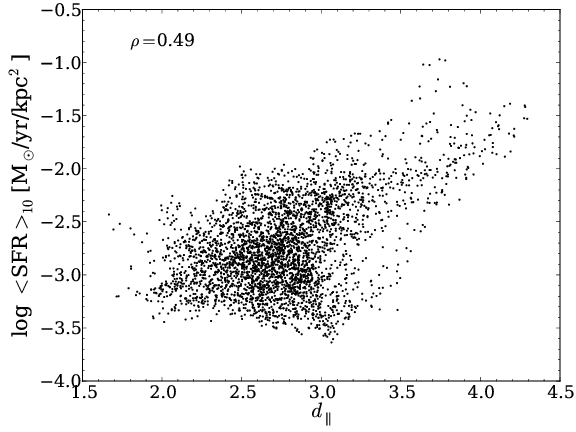}
\includegraphics[width=0.6\columnwidth]{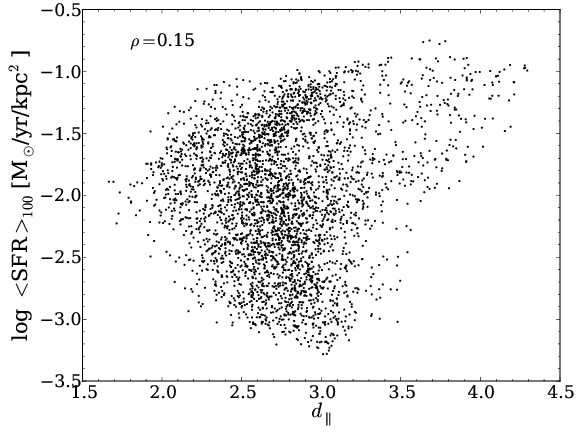}
\includegraphics[width=0.6\columnwidth]{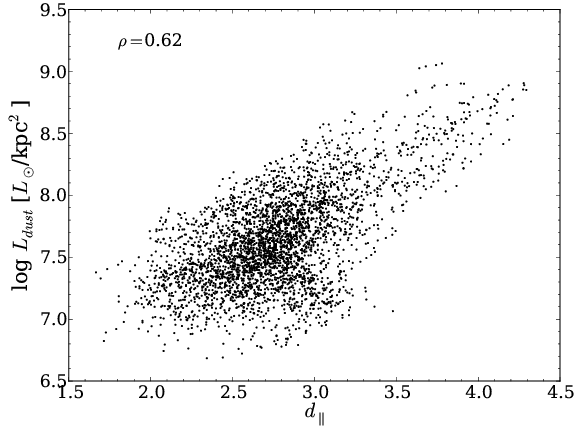}
\includegraphics[width=0.6\columnwidth]{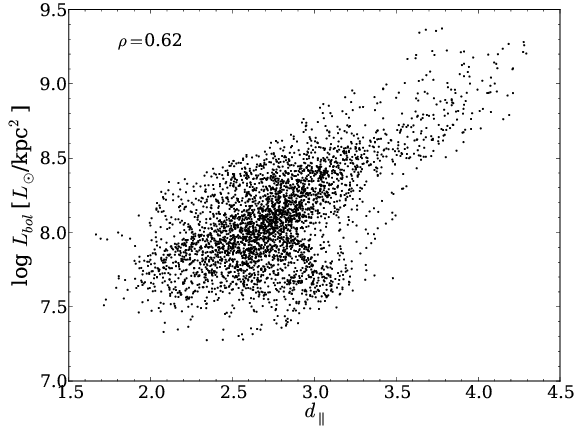}
\includegraphics[width=0.6\columnwidth]{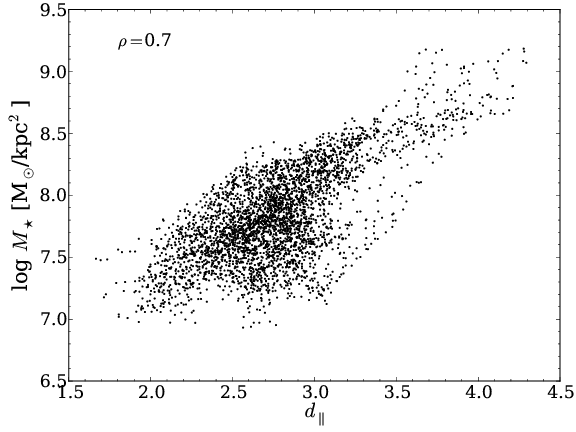}
\includegraphics[width=0.6\columnwidth]{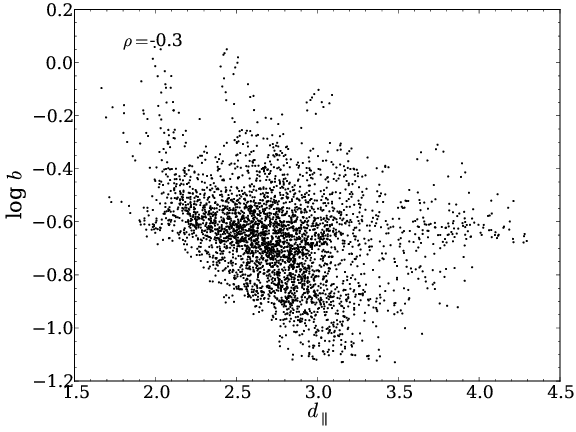}
\includegraphics[width=0.6\columnwidth]{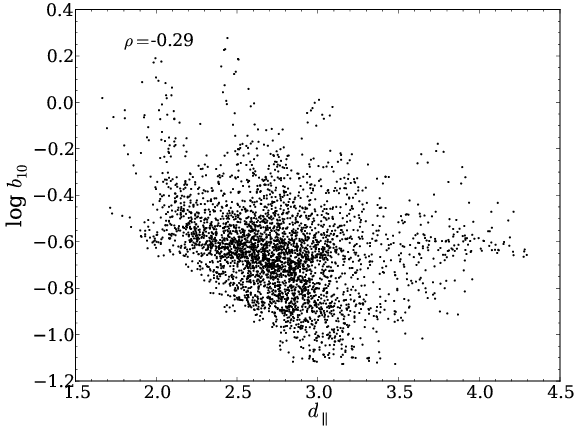}
\includegraphics[width=0.6\columnwidth]{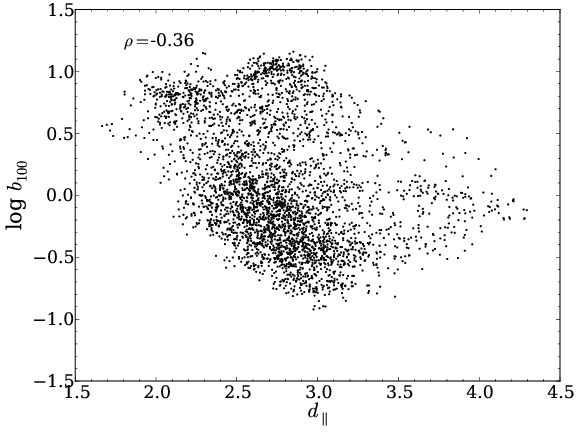}
\caption{Parameters versus $d_\parallel$. The correlation coefficient is indicated on the top--left hand corner of each plot.}
\label{fig:dist-para}
\end{figure*}

\end{document}